\documentclass[showpacs,prl,onecolumn,aps,superscriptaddress,preprintnumbers,nofootinbib]{revtex4}
\usepackage[T1]{fontenc}
\usepackage[latin9]{inputenc}
\usepackage{amsmath,amssymb}
\usepackage{epsfig}
\usepackage{graphicx}
\usepackage{mathrsfs}
\usepackage{amsmath}
\usepackage{amsfonts}
\usepackage{epstopdf}
\usepackage{color}
\def\slashchar#1{\setbox0=\hbox{$#1$}     		% set a box for #1
   \dimen0=\wd0                                 	% and get its size
   \setbox1=\hbox{/} \dimen1=\wd1               	% get size of /
   \ifdim\dimen0>\dimen1                        	% #1 is bigger
      \rlap{\hbox to \dimen0{\hfil/\hfil}}      	% so center / in box
      #1                                        	% and print #1
   \else                                        	% / is bigger
      \rlap{\hbox to \dimen1{\hfil$#1$\hfil}}   	% so center #1
      /                                         	% and print /
   \fi}

\renewcommand{\vec}{\boldsymbol}
\newcommand{\be}{\begin{equation}}
\newcommand{\ee}{\end{equation}}
\newcommand{\bea}{\begin{eqnarray}}
\newcommand{\eea}{\end{eqnarray}}
\newcommand{\ba}{\begin{array}}
\newcommand{\ea}{\end{array}}

\def\eq#1{{Eq.~(\ref{#1})}}
\def\fig#1{{Fig.~\ref{#1}}}
\newcommand{\bas}{\bar{\alpha}_S}

\newcommand{\nn}{\nonumber}

\newcommand{\Lb}{\left(}
\newcommand{\Rb}{\right)}
\newcommand{\h}{\frac{1}{2}}
\def\pom{{I\!\!P}}

\begin{document}

%\title{Parton densities without partons}
\title{ High energy QCD: multiplicity dependence of quarkonia production}
\author{E. ~Gotsman}
\email{gotsman@post.tau.ac.il}
\affiliation{Department of Particle Physics, School of Physics and Astronomy,
Raymond and Beverly Sackler
 Faculty of Exact Science, Tel Aviv University, Tel Aviv, 69978, Israel}
 \author{E.~ Levin}
\email{leving@tauex.tau.ac.il, eugeny.levin@usm.cl}
\affiliation{Department of Particle Physics, School of Physics and Astronomy,
Raymond and Beverly Sackler
 Faculty of Exact Science, Tel Aviv University, Tel Aviv, 69978, Israel}\affiliation{ Departamento de F\'\i sica,
Universidad T$\acute{e}$cnica Federico Santa Mar\'\i a   and
Centro Cient\'\i fico-Tecnol$\acute{o}$gico de Valpara\'\i so,
Casilla 110-V,  Valparaiso, Chile}

\date{\today}


\begin{abstract}

In this paper we   propose an approach which demonstrates  the  
dependence of quarkonia
 production on the multiplicity of the accompanying hadrons. Our approach  
is 
based on
 the three gluon fusion mechanism, without  assuming the multiplicity
 dependence of the saturation scale. We show, that we  describe the
 experimental data, which   has a dependence that is much
 steeper than the multiplicity of the hadrons.

\end{abstract}
\maketitle

\vspace{-0.5cm}
\tableofcontents

%%%%%%%%%%%%%%%%%%%%%%%%%%%%%%%%%%%%%%%%%%%%%%%%%%%%
\section{Introduction}
The goal of this paper is to study the multiplicity dependence of 
 quarkonia (mainly $J/\Psi$)  production in  the framework of high
 energy QCD 
 (see Ref.\cite{KOLEB} for a general review).  Effective  QCD at
 high energies currently exists in two different formulations:
   the CGC/saturation approach
 \cite{MV,MUCD,B,K,JIMWLK,GIJMV}, and the BFKL Pomeron calculus 
\cite{BFKL,LI,GLR,GLR1,MUQI,MUPA,BART,BRN,KOLE,LELU1,LELU2,LMP,
AKLL,AKLL1,LEPP}.  In 
  this paper we restrict ourself   to  the BFKL 
Pomeron
 calculus, which has a more direct correspondence with the parton 
approach, and which  provides  an approximation for  estimates of 
hadron-hadron
 collisions, that at present are out of the reach   for  the CGC 
approach. 
  
  Fortunately, in Ref.\cite{AKLL1} it was shown, that 
these
 two approaches are equivalent for the description of the scattering
 amplitude in the rapidity range :
\be \label{I2}
Y \,\leq\,\frac{2}{\Delta_{\mbox{\tiny BFKL}}}\,\ln\Lb
 \frac{1}{\Delta^2_{\mbox
{\tiny BFKL}}}\Rb
\ee
where $\Delta_{\mbox{\tiny BFKL}}$ denotes the intercept of the BFKL 
 Pomeron. In this paper it is also shown, that for \eq{I2} we can use the
 Mueller, Patel, Salam and Iancu approximation(MPSI) \cite{MPSI} for
 hadron-hadron scattering at high energies.

\begin{figure}[ht]
     \begin{center}
     \includegraphics[width=0.6\textwidth]{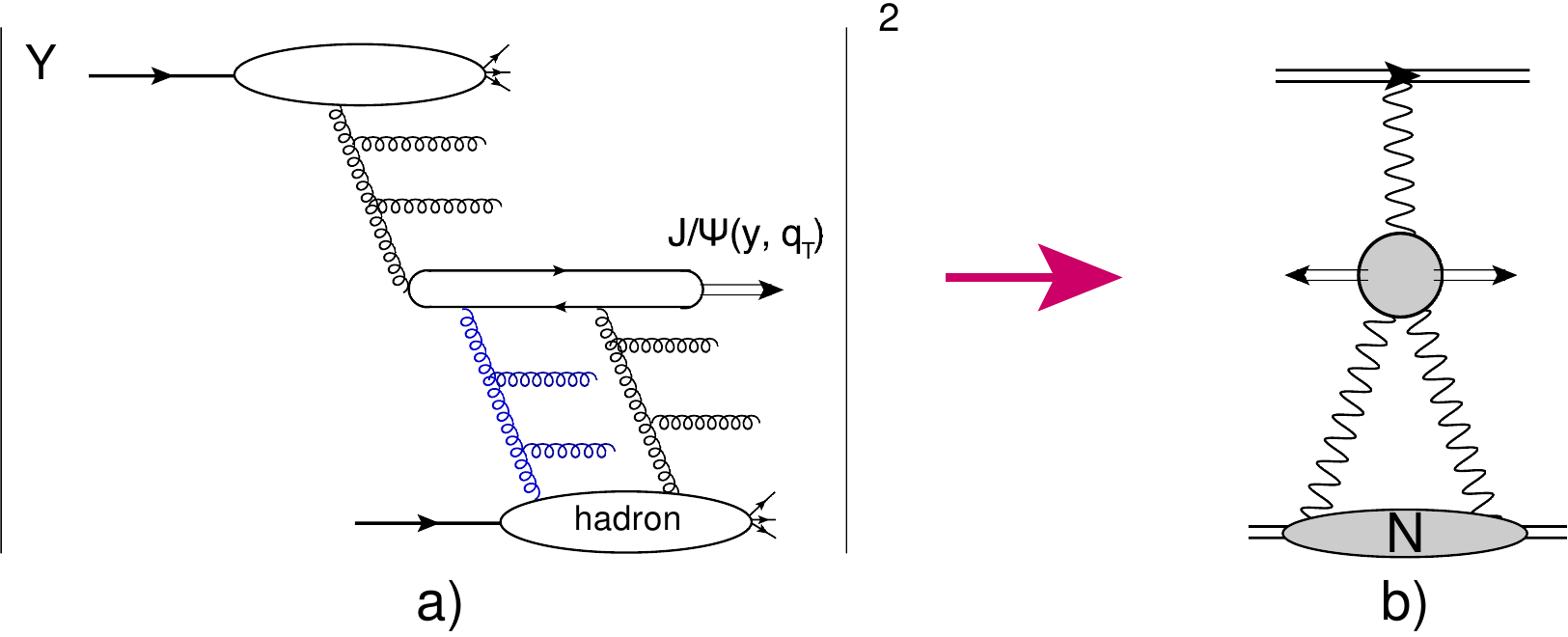} 
     \end{center}    
      \caption{ \fig{3p}-a: The three gluon fusion mechanism
 of $J/\Psi$ production.\fig{3p}-b: the Mueller diagram\cite{MUDIA}
 for $J/\Psi$ production, which illustrates the inter-relation between
 three gluon fusion and the triple BFKL Pomeron interaction. The wavy 
lines
 describe the BFKL Pomerons, while the helical curves represent 
gluons.}
\label{3p}
   \end{figure}

The recent experiments by  ALICE\cite{ALICE0,ALICE1,ALICE01,ALICE2,ALICE3} 
 and STAR\cite{STAR1,STAR2}, show that the cross sections for $J/\Psi$
 production    depends strongly  on the multiplicity of accompanying 
hadrons.
 These data have  stimulated theoretical discussions on the origin of such
 dependence (see Refs.\cite{KPPRS,FEPA,MTVW,LESI,LSS}). In this paper, we
 develop an approach to this problem based on two ingredients. First, we
 assume ,that the production of quarkonia stems from  triple gluon
 fusion\cite{KMRS,MOSA,LESI} (see \fig{3p}).  For the interaction with
 nuclei this mechanism is dominant \cite{KHTU,KLNT,DKLMT,KLTPSI,KMV}; and
it has been demonstrated
 in Ref.\cite{LESI},  that this mechanism gives a
 substantial contribution in hadron-hadron collisions.

 %%%%%%%%%%%%%%%%%%%%%%%%%%%%%%%%%%%%%%%%%%%%%%%%%%%%%%%%%%%%%%%
%%%%%%%%%%%%%%%%%%%%%%%%%%%%%%%%%%%%%%%%%%%%%%%%%%%%%%%
Second, we showed in Refs.\cite{GOLEMULT,KHLE}, that in spite of the fact
 that in different kinematic
 regions,  the QCD cascade leads to a different energy and dipole size
 dependence of the mean multiplicity, the multiplicity distribution 
has
 a general form:
\be \label{I3}
\frac{\sigma_n}{\sigma_{ \rm in}}\,\,=\,\,\frac{1}{N}\,\Lb \frac{N\,-\,1}{N}\Rb^{n - 1}\,
\ee
 where N denotes the
 average number of partons. 
 
 The paper is organized as follows: In the next section we describe our
 approach to hadron-hadron collisions. In section III we  discuss
 quarkonia production in  simplified  Reggeon Field Theory, defined
 in zero transverse dimensions.  In section IV we generalize  the result
 in this toy-model approach to high energy QCD, and compare our estimates
 with the experimental data .
We summarize our results in the Conclusions.

%%%%%%%%%%%%%%%%%%%%%%%%%%%%%%%%%%%%%%%%%%%%%%%%%%%%
\section{Hadron-hadron interaction in MPSI approach }
%%%%%%%%%%%%%%%%%%%%%%%%%%%%%%%%%%%%%%%%%%%%%%%%%%%%%
This section does not contain  new results, and we include it in the paper 
for
  completeness of presentation, as well as a kind of an introduction to
 the notation, and the  main ideas.

   %%%%%%%%%%%%%%%%%%%%%%%%%%%%%%%%%%%%%%%%%%%%%%%%%%
 \subsection{QCD parton cascade}
 %%%%%%%%%%%%%%%%%%%%%%%%%%%%%%%%%%%%%%%%%%%%%%%%
 
 We start with the equation for the QCD parton cascade which   can
 be written  in the following form.\cite{KOLEB,MUCD,LELU1,LELU2}:
  \bea  \label{PC1}
&&\frac{\partial\,P_n\left(Y, \vec{r }, \vec{b};\,\vec{r}_1,\vec{ b}_1,\,\vec{r}_2 , \vec{b}_2\dots \vec{r}_i ,\vec{b}_i,
\dots \vec{r}_n, \vec{b}_n \right)}{ 
\partial\, Y }\,=\,-\,
\sum^n_{i=1}\,\omega_G(r_i) \,
P_n\left(Y, \vec{r }, \vec{b};\,\vec{r}_1,\vec{ b}_1,\,\vec{r}_2 , \vec{b}_2\dots \vec{r}_i ,\vec{b}_i,
\dots \vec{r}_n, \vec{b}_n \right) \\
&&~~~~~~~~~~~~~~~~~~~~~~~~~~~~~~~~~~~~~~~~~~~~~+\,\,\bas\,\sum^{n-1}_{i=1} \,\frac{(\vec{r}_i\,+\, 
\vec{r}_n)^2}{(2\,\pi)\,r^2_i\,r^2_n}\,
P_{n - 1}\left(Y, \vec{r},\vec{b};\,\vec{r}_1,\vec{b}_1,
\dots  (\vec{r}_i \,+\, \vec{r}_n), \vec{b}_{in},\dots \vec{r}_{n-1},\vec{b}_n
 \right)\nn
\eea
  where $P_n\Lb Y ; \{r_i,b_i\}\Rb$ denotes the probability to have 
$n$-dipoles
 of size $r_i$,  at impact parameter $b_i$, and  at rapidity $Y$\footnote
{ In the lab. frame rapidity $Y$ is equal to   $Y = y_{\rm dipole ~r} \,-
\,y_{\rm dipoles~r_i}$, where $y_{\rm dipole ~r}$ is the rapidity of
 the incoming fast dipole and $y_{\rm dipole~r_i}$ is the rapidity of
 dipoles $r_i$.}
  . $\vec{b}_{in} $ in \eq{PC1} is given by $\vec{b}_{in}
 \,=\,\vec{b}_i \,+\,\h \vec{r}_i \,=\,\vec{b}_n \,-\,\h \vec{r}_i$.
  
  \eq{PC1} is a typical cascade equation in which the first term
 describes  the depletion of the probability  of   $n$, due to  one dipole 
   decaying into two dipoles 
of  arbitrary sizes, while the second term describes,  the growth due to 
the 
splitting
 of ($n - 1$) dipoles into
$n$ dipoles.

The initial condition for the DIS scattering is
\be \label{PCIC}
P_1 \Lb Y  =  0,  \vec{r},\vec{b} ; \vec{r}_1,\vec{b}_1\Rb\,\, =\,\,\,\delta^{(2)}\Lb \vec{r}\,-\,\vec{r}_1\Rb\,\delta^{(2)}\Lb \vec{b}\,-\,\vec{b}_1\Rb;~~~~~~~P_{n>1}\Lb Y  = 0; \{r_i\}\Rb \,=\,0
\ee
which corresponds to the fact that we are discussing a dipole of 
 definite size which develops the parton cascade.

Since $P_n\Lb Y ; \{r_i\}\Rb$ is the probability to find dipoles $\{r_i\}$,
 we have the following sum rule 

\be \label{SUMRU}
\sum_{n=1}^\infty\,\int \prod^n_{i=1} d^2 r_i \,d^2 b_i \,P_n\Lb Y ; \{\vec{r}_i\,\vec{b}_i\}\Rb\,\,=\,\,1 ,
\ee
i.e. the sum of all probabilities is equal to 1.

 This QCD  cascade leads to the Balitsky-Kovchegov (BK) equation 
\cite{B,K,KOLEB} for the  amplitude, and gives the theoretical
 description of  DIS.  To see this  we introduce the generating
 functional\cite{MUCD}

\be \label{Z}
Z\Lb Y, \vec{r},\vec{b}; [u_i]\Rb\,\,=\,\,\sum^{\infty}_{n=1}\int P_n\Lb Y,\vec{r},\vec{b};\{\vec{r}_i\,\vec{b}_i\}\Rb \prod^{n}_{i=1} u\Lb \vec{r}_i\,\vec{b}_i\Rb\,d^2 r_i\,d^2 b_i
\ee
 where $u\Lb \vec{r}_i\,\vec{b}_i\Rb \equiv\,=u_i$ is an arbitrary function.
 The initial conditions of \eq{PCIC}  and the sum rules of \eq{SUMRU} 
require the 
 following form for the functional $Z$:
\begin{subequations}
\bea
Z\Lb Y=0, \vec{r},\vec{b}; [u_i]\Rb &\,\,=\,\,&u\Lb \vec{r},\vec{b}\Rb;\label{ZIC}\\
Z\Lb Y, r,[u_i=1]\Rb &=& 1; \label{ZSR}
\eea
\end{subequations}

Multiplying both terms of \eq{PC1} by $\prod^{n}_{i=1} u\Lb \vec{r}_i
\,\vec{b}_i\Rb$ and integrating over $r_i$ and $b_i$, we obtain the
 following linear functional equation\cite{LELU2};
\begin{subequations}
\bea
&&\hspace{-0.7cm}\frac{\partial Z\Lb Y, \vec{r},\vec{b}; [u_i]\Rb}{\partial \,Y} =\int d^2 r'\,  K\Lb \vec{r}',\vec{r} - \vec{r'}|\vec{r}\Rb\Bigg( - u\Lb r, b\Rb\,\,+\,\,u\Lb \vec{r}',\vec{b} + \h(\vec{r} - \vec{r}') \Rb \,u\Lb \vec{r} - \vec{r}',\vec{b} - \h\vec{r}'\Rb\Bigg) \frac{\delta\,Z}{\delta \,u\Lb r, b \Rb};\label{EQZ}\\
&&  K\Lb \vec{r}',\vec{r} - \vec{r'}|\vec{r}\Rb\,=\frac{\bas}{2 \,\pi}\frac{r^2}{r'^2\,(\vec{r} - \vec{r}')^2} ;\,~~~~~
 \omega_G\Lb r\Rb\,\,=\,\,\int d^2 r'  K\Lb \vec{r}',\vec{r} - \vec{r'}|\vec{r}\Rb; \label{OMG}
 \eea
\end{subequations}
Searching for a solution of the form     $Z\Lb [ u(r_i,b_i,Y)]\Rb$  
for the initial conditions of \eq{ZIC}, \eq{EQZ} can be re-written as
 the non-linear equation \cite{MUCD}:
\be \label{NEQZ}
\frac{\partial Z\Lb Y, \vec{r},\vec{b}; [u_i]\Rb}{\partial \,Y}\,=\,\int d^2 r' K\Lb \vec{r}',\vec{r} - \vec{r'}|\vec{r}\Rb\Bigg\{Z\Lb r',  \vec{b} + \h(\vec{r} - \vec{r}'); [u_i]\Rb  \,Z\Lb \vec{r} - \vec{r'},  \vec{b} - \h\vec{r}'; [u_i]\Rb
\,\,-\,\,Z\Lb Y, \vec{r},\vec{b}; [u_i]\Rb\Bigg\}
\ee
Therefore, the QCD parton cascade of \eq{PC1} takes into account 
 non-linear evolution. 
Generally speaking the scattering amplitude can be written in the
 form\cite{K,LELU2}:
\be \label{N1}
N(Y,\,r,\,b)\,=\,-\,\,\sum^{\infty}_{n=1}\,(-1)^n\,
\rho^p_n(r_1,\, b_1,\,\ldots\,r_n,\,b_n\,;\,Y\,-\,Y_0)
\,\,\prod^n_{i =1}\,N(Y_0,\,r_i,\,b_i)\,\,
d^2\, r_i\, \, d^2\, b_i \,.
\ee

where $N(Y_0,\,r_i,\,b_i)$ is the amplitude of the interaction of dipole
 $r_i$ with the target at low energy $Y=Y_0$, and the $n$-dipole densities
 in the projectile 
$\rho^p_n(r_1, b_1,\ldots\,,r_n, b_n)$
 are defined as follows:
\be \label{N2}
\rho^p_n(r_1, b_1\,
\ldots\,,r_n, b_n; Y\,-\,Y_0)\,=\,\frac{1}{n!}\,\prod^n_{i =1}
\,\frac{\delta}{\delta
u_i } \,Z\left(Y\,-\,Y_0;\,[u] \right)|_{u=1}
\ee
For $\rho_n$ we obtain\cite{LELU2} :
\bea \label{N3}
\frac{\partial \,\rho^p_n(r_1, b_1\,\ldots\,,r_n, b_n)}{ 
\bar{\alpha}_s\,\partial\,Y}\,\,&=&\,\
-\,\sum_{i=1}^n
 \,\,\omega(r_i)\,\,\rho^p_n(r_1, b_1\,\ldots\,,r_n, b_n)\,\,+\,\,2\,\sum_{i=1}^n\,
\int\,\frac{d^2\,r'}{2\,\pi}\,
\frac{r'^2}{r^2_i\,(\vec{r}_i\,-\,\vec{r}')^2}\,
\rho^p_n(\ldots\,r', b_i-r'/2\dots)\nn\\
 & & 
\,
+\,\sum_{i=1}^{n-1}\,\frac{(\vec{r}_i + \vec{r}_n)^2}
{(2\,\pi)\,r^2_i\,r^2_n}\,
\rho^p_{n-1}(\ldots\,(\vec{r}_i\,+\,\vec{r}_n), b_{in}\dots).
\eea
For $\rho_1$ 
we have  the linear  BFKL equation\cite{BFKL}: 
\be \label{N4}
\frac{\partial \,\rho^p_1(Y; r_1, b)}{ 
\bas\,\partial\,Y}\,\,=\,\,-\,\,\omega_G\Lb r_1\Rb\rho^p_1(Y; r_1, b) \,\,+\,2\,\int\,\frac{d^2\,r'}{2\,\pi}\,
\frac{r'^2}{r^2_1\,(\vec{r}_1\,-\,\vec{r}')^2}\,
\bar{\rho}^p_1\Lb Y, r',b\Rb
\ee
However, to obtain the BK equation for the 
scattering amplitude we need to use \eq{N1}, in which we introduce
 the amplitude of interaction of the dipole with the target at low
 energies. Using  \eq{EQZ},\eq{N1}  and  \eq{N2} , we can obtain the
 non-linear BK equation from \eq{NEQZ} in the following form\cite{K}
\bea 
\frac{\partial}{\partial Y} N\Lb \vec{r}, \vec{b} ,  Y \Rb &=&\int d^2 r'\,K\Lb \vec{r}', \vec{r} - \vec{r}'| \vec{r}\Rb \Bigg\{N\Lb \vec{r}',\vec{b} - \h \Lb \vec{r} - \vec{r}' \Rb, Y\Rb + 
N\Lb\vec{r} - \vec{r}', \vec{b} - \h \vec{r}', Y\Rb \,\,- \,\,N\Lb \vec{r},\vec{b},Y \Rb\nn\\
& & ~~~~~~~- N\Lb\vec{r} - \vec{r}', \vec{b} - \h \vec{r}', Y\Rb   \, N\Lb \vec{r}',\vec{b} - \h \Lb \vec{r} - \vec{r}' \Rb, Y\Rb\Bigg\}\label{NEEQa}
\eea
   %%%%%%%%%%%%%%%%%%%%%%%%%%%%%%%%%%%%%%%%%%%%%%%%%%
 \subsection{The interaction of two dipoles at high energies}
\begin{figure}[ht]
     \begin{center}
     \begin{tabular}{c c c}
     \includegraphics[width= 0.5\textwidth]{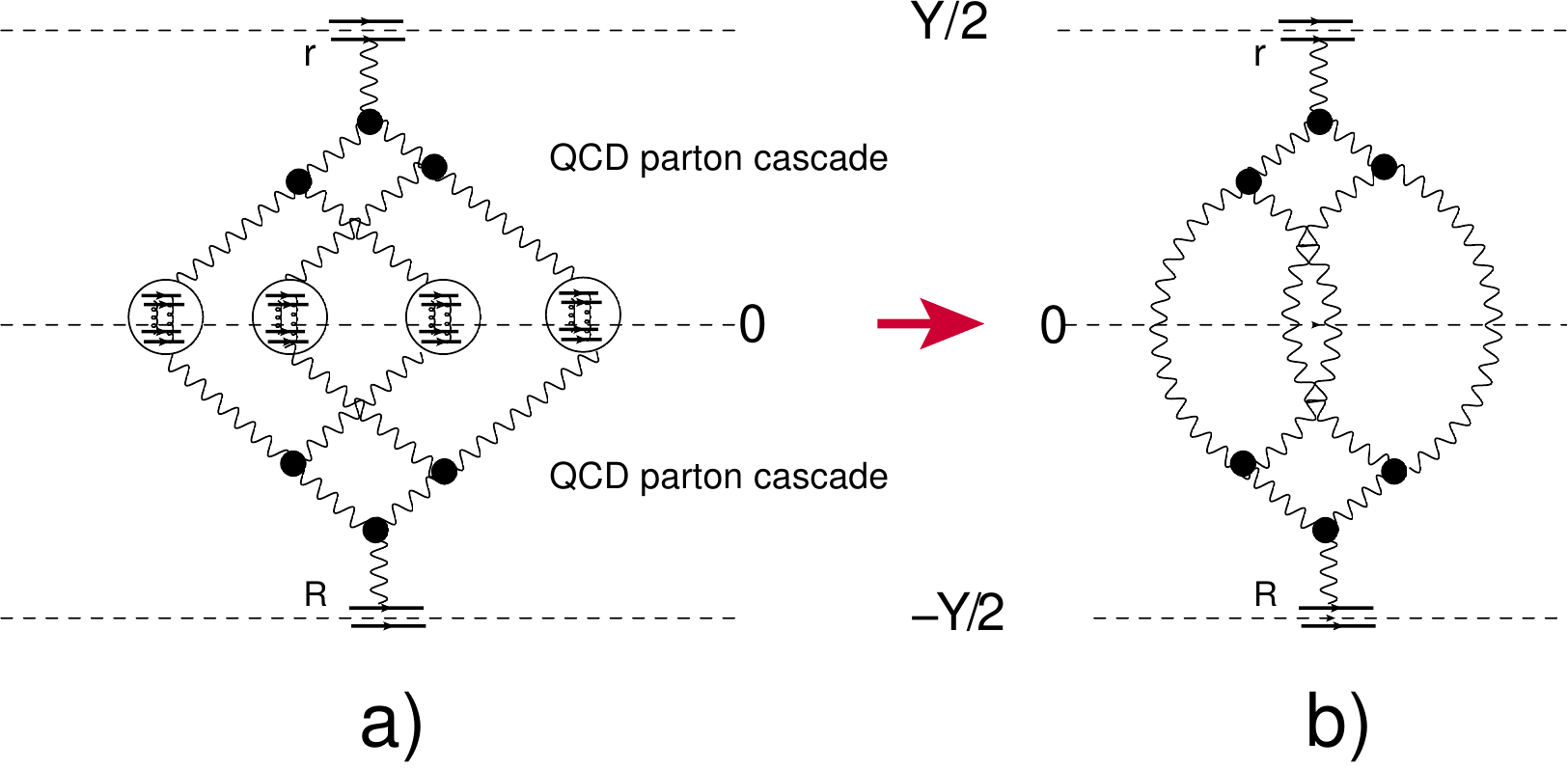} &~~~~~~~~~& \includegraphics[width=0.21\textwidth]{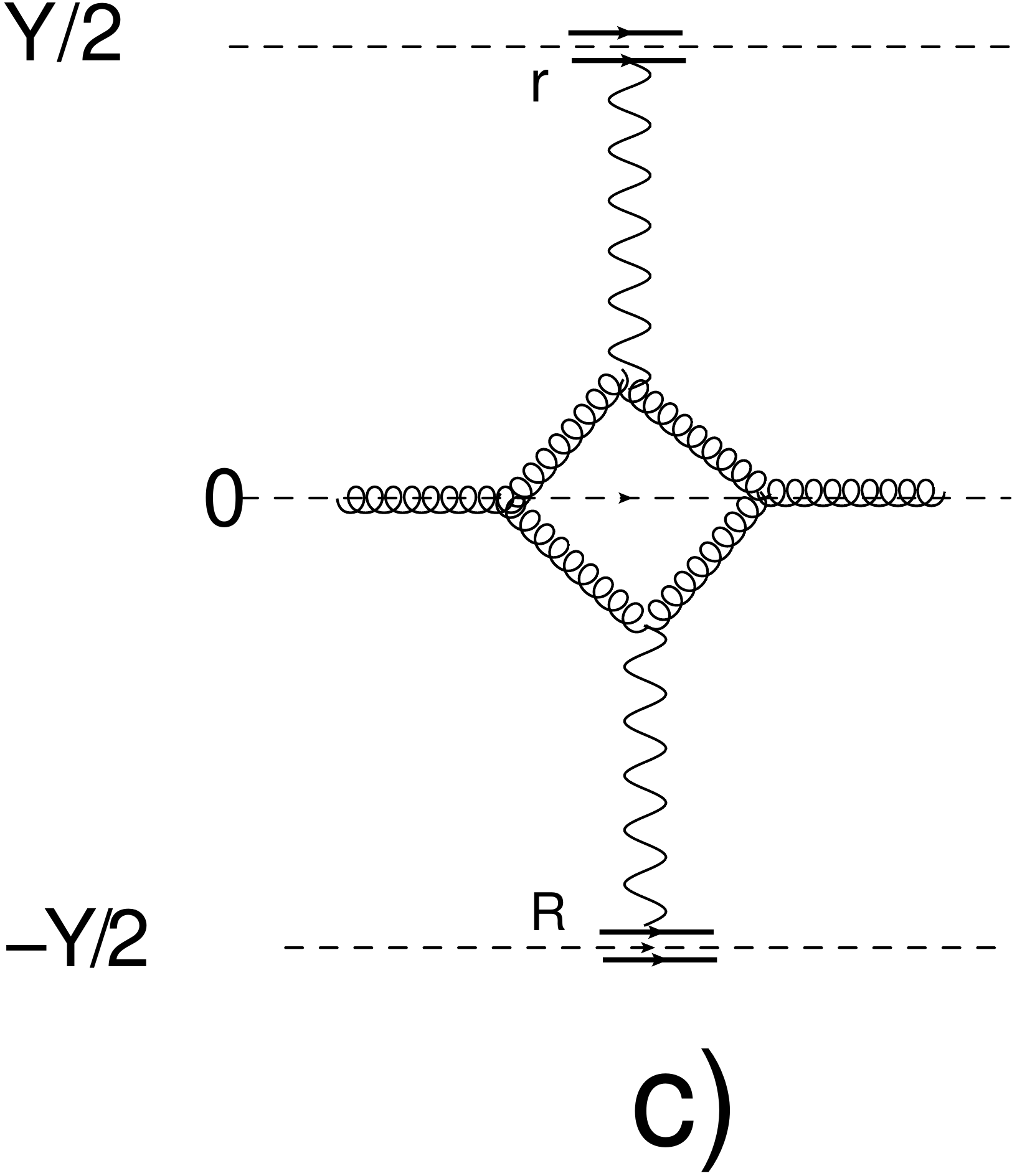}\\
    \end{tabular}     \end{center}    
      \caption{ Scattering amplitude for the interaction of two dipoles
 with sizes:  $r$ and $R$ at high energy in the  MPSI approach
 (see \fig{mpsi}-a and \fig{mpsi}-b). The amplitudes of
 interaction of two dipoles in the Born approximation of
 perturbative QCD ( $N\Lb r_i,r'_i,b"_i\Rb$ in \eq{HH1})
 are shown   as  white circles. The wavy  lines denote 
the BFKL Pomerons. \fig{mpsi}-c shows the Mueller diagram\cite{MUDIA}
 for inclusive production of gluons. }
\label{mpsi}
   \end{figure}

 %%%%%%%%%%%%%%%%%%%%%%%%%%%%%%%%%%%%%%%%%%%%%%%% 
 We first  consider  the simplest  case of scattering, the high energy 
interactions of two dipoles  
 with sizes $r$ and $R$ and with $r \,\sim\, R$.
 In Ref.\cite{AKLL1} it is shown that in the limited range of rapidities,
  given by \eq{I2}, we can safely apply 
 the  Mueller, Patel, Salam and Iancu approach for this 
scattering \cite{MPSI}(see
 \fig{mpsi}-a).  
 
  %%%%%%%%%%%%%%%%%%%%%%%%%%%%%%%%%%%%%%%%%%%%%%%%%%%%%%%%%%%%%%%
%%%%%%%%%%%%%%%%%%%%%%%%%%%%%%%%%%%%%%%%%%%%%%%%%%%%%%% 

The scattering amplitude in this approach can be written in the following
 form\cite{LELU2}:

\bea \label{HH1}
 N\Lb Y, r, R, b \Rb \,\,&=&\,\,-\,\, \sum^{\infty}_{n =1}\,\,(-1)^n\,
\int\,\,\rho^t_n\Lb \vec{r}_1, \,\vec{b}'_1 ,\ldots \,, \vec{r}_n,\,  \vec{b}'_n;\,\h Y\Rb\,\,
\rho^p_n\Lb \vec{r}'_1, \, \vec{b} - \vec{b}'_1 - \vec{b}_1''\,,\ldots\,,\vec{r}_n,\,  \vec{b} - \vec{b}'_n - \vec{b}_n'';\, -\h Y\Rb \nn\\
& \times& \,\prod^n_{i =1}\, \,
d^2\, r_i \,
\,\,\prod^n_{j =1}\, \,d^2 \,r'_j \, d^2 b'_j d^2 b''_j\,N^{\rm BA}\Lb r_i, r'_i,b''_{i}\Rb
\eea

where $\rho^t_n$ and     $\rho^p_n$ denote the parton densities in the 
target 
and projectile, respectively. These densities can be calculated from $P_n$ 
using \eq{N2}. $N^{\rm BA} $ is the scattering amplitude of two dipoles in
 the Born approximation of perturbative QCD (see \fig{mpsi}).  
  \eq{HH1} simply  states that we can consider the QCD parton cascade of
 \eq{PC1} generated by the dipole of  size $r$ for 
the c.m.f.  rapidities from 0 to $\h Y$, and the same cascade for 
the dipole
 of the size $R$, for the rapidities from 0 to $ -\h Y  $.  One can see
 that \eq{HH1} is the $t$-channel unitarity re-written in a form,
 convenient for applying the evolution of the parton cascade in the
  form  of \eq{N3}.

 Generally speaking, for a dense system of partons at $Y \, =\,0$
 $n$-dipoles from upper cascade could  interact with $m$ dipoles 
from the lower cascade, with the amplitude $N^m_n\Lb\{r_i\},\{r'_j\}\Rb$
\cite{LELU2}.
 In \eq{HH1} we assume that the system of dipoles that has been created
 at $Y =0$ is not very dense, at least for the range of rapidities given
 by \eq{I2}. In this case 
 \be \label{HH10}
 N^m_n\Lb\{r_i\},\{r'_j\}\Rb\,\,=\,\,\delta_{n,m}\prod^n_{j =1}\,\Lb - 1\Rb^{n - 1}N^{\rm BA}\Lb r_i, r'_i,b''_i\Rb
\ee

 and after integration over $\{r_1\}$ and $\{r'_j\}$, the scattering
 amplitude can be reduced to a system of enhanced BFKL Pomeron
 diagrams, which are shown in \fig{mpsi}-b.
 
 The average number of dipoles at $Y = 0$ is determined by the 
inclusive cross section, which is given by the diagram of
 \fig{mpsi}-c and which can be written at $y \to 0$  as
 follows\cite{KTINC}:
 \be \label{HH2}
\frac{d \sigma}{d y \,d^2 p_{T}} = \frac{2C_F}{\alpha_s (2\pi)^4}\,\frac{1}{p^2_T} \int d^2 \vec r_T\,e^{i \vec{p}_T\cdot \vec{r}_T}\!\!\int\!\! d^2 b\,\nabla^2_T\,N^{\rm \mbox{\tiny  BFKL}}\Lb \h Y ;r,  r_T; b \Rb\,\int \!\!d^2 B \,\nabla^2_T\,N^{\rm\mbox{\tiny  BFKL}}\Lb y_2 = -\h Y ;R,  r_T; B \Rb
\ee
The average number of dipoles that enters the multiplicity
 distribution of \eq{NDIST}, is equal $ \bar{n} \,=\,N\,=\int
 \frac{d^2 p_T}{(2 \pi)^2} \frac{d \sigma}{d y \,d^2 p_{T}}\Big{/}
 \sigma_{in} \,\propto\,\exp\Lb \Delta_{\rm BFKL}\,Y\Rb$\footnote{$
 \Delta_{\rm BFKL}$ denotes the intercept of the BFKL Pomeron.}
only if we assume that $\sigma_{in} \,\sim {\rm Const}$. 
Indeed, the enhanced diagrams of \fig{mpsi}-b lead to the
 inelastic cross section which is constant at high energy.

   %%%%%%%%%%%%%%%%%%%%%%%%%%%%%%%%%%%%%%%%%%%%%%%%%%
 \subsection{Hadron - hadron collisions}
 %%%%%%%%%%%%%%%%%%%%%%%%%%%%%%%%%%%%%%%%%%%%%%%%

In this paper we  
view a  hadron as a dilute system of dipoles and 
use
 \eq{HH2} for the average multiplicity, 
   together with the multiplicity  distribution of \eq{I2}. In particular,
 we assume that \eq{HH10} is correct, and the system of partons that is
 produced at c.m. rapidity $y^*$=0 is a dilute system. However, we are
 aware that \eq{HH2} does not describe the experimental increase of
 the average multiplicity,  which from \eq{HH2} is
    $\bar{n} \,\propto\,\exp\Lb \Delta_{\rm BFKL}\,Y\Rb$.   The experimental
 data can be described in the
 framework of the CGC/saturation approach in which $N^{\rm 
\mbox{\tiny  BFKL}}$ were replaced by $N^{\rm \mbox{\tiny BK}}$\cite{LERE}.
Hence, we cannot view hadrons as a dilute system of 
 dipoles, but rather have to consider them as 
a dense system of dipoles.  For such a situation we expect
 that $\bar{n} \,\propto \,Q^2_s(Y)/\bas$ (see Refs.\cite{KOLEB,KLN,DKLN,
LERE,LAPPI}.     
   
%%%%%%%%%%%%%%%%%%%%%%%%%%%%%%%%%%%%%%%%%%%
\section{Reggeon Field Theory  in zero transverse dimensions.}
%%%%%%%%%%%%%%%%%%%%%%%%%%%%%%%%%%%%%%%%%%%%%
   
%%%%%%%%%%%%%%%%%%%%%%%%%%%%%%%%%%%%%%%%%%%
\subsection{Multiplicity distribution - a recap}
%%%%%%%%%%%%%%%%%%%%%%%%%%%%%%%%%%%%%%%%%%%%
%%%%%%%%%%%%%%%%%%%%%%%%%%%%%%%%%%%%%%%%%%%%%%%%%%%%
In the parton model\cite{FEYN,BJP,Gribov} all partons have average transverse
 momentum 
which does not depend
 on energy. Therefore, we can obtain the parton model from the QCD
 cascade assuming that the unknown confinement of gluons leads to 
the QCD cascade for a dipole of fixed size. In this case the
 cascade equation (see \eq{PC1})  takes the following simple form:
\be \label{EQPM}
 \frac{d P_n\Lb Y\Rb}{d Y}\,\,=\,\,- \Delta\,n\,  P_n\Lb Y\Rb \,\,+\,\,\Lb n - 1 \Rb \Delta P_{n-1}\Lb Y .\Rb
\ee
where  $P_n\Lb Y \Rb$ denotes the probability to find $n$ dipoles
 (of a fixed size in our model) at  rapidity $Y$, and  $\Delta$ denotes
 the intercept of the BFKL Pomeron. 

Instead of the generating functional of \eq{Z}, we can introduce the
 generating function:

\be \label{ZF}
Z\Lb Y, u\Rb\,\,=\,\,\sum_n \,P_n\Lb Y\Rb \,u^n 
\ee
where $u$ are  numbers.

At the initial rapidity $Y=0$, we have only one dipole,  so $P_1\Lb Y = 0
 \Rb =1$ and $P_{n > 1} \,=\,0$ (so the state is  only one dipole);
  at $u = 1$,  $Z\Lb  Y, u=1\Rb\,\,=\,\,\sum_n P\Lb y \Rb\,=\,1$. These
 two properties  determine the initial and the boundary conditions for
 the generating function which 	simplify \eq{ZIC} and \eq{ZSR}
\be \label{TM3}
Z\Lb Y = 0,u\Rb \,\,=\,\,u;~~~~~~~~~Z\Lb Y, u=1\Rb\,\,=\,\,1.
\ee
 \eq{EQPM} takes the  following form for the generating function:
\be \label{TM4}
\frac{\partial  Z\Lb Y, u\Rb}{ \partial Y}\,\,=\,\,- \Delta \, u \Lb 1 - u\Rb \frac{\partial Z\Lb Y, u\Rb}{\partial u} .
\ee
%%%%%

 The general solution to \eq{TM4} is an arbitrary function 
 ($  Z\Lb z \Rb$) of the new variable: $z\,\,=\,\,\Delta\,Y\,\,+\,\,f(u)$, 
with
  f(u)  from the following equation:
\be \label{TM5}
1\,\,=\,\,-\,u\,\Lb 1\,-\,u\Rb\,f'_u\Lb u\Rb  ~~~~~f\Lb u\Rb\,\,=\,\,\ln\Lb \frac{u\,-\,1}{u}\Rb\,\,+\,\,C_1
\ee
The form of arbitrary function stems from the initial condition of \eq{TM3} 
\be \label{TM6}
Z\Lb z\Lb Y=0\Rb\Rb\,=\,u;  
\ee
Since $u\,\,=\,\,1\Big{/}\Lb 1\,-\,e^z\Rb$ we obtain that
\be \label{SOLTM}
Z\Lb Y,\,u\Rb\,\,=\,\,\frac{u\,\,e^{\,-\,\Delta\, Y}}{1\,\,+\,\,u\,\,
\Lb e^{\,-\,\Delta\,Y}\,-\,1\Rb}\,\,=\,\,u\,e^{\,-\,\Delta\, Y}\,\sum^\infty_{n=1}\,u^n \Lb 1\,-\, e^{\,-\,\Delta\,Y}\Rb^{n \,-\,1} .
\ee
Note, that $Z\Lb Y,u=1\Rb = 1$, as it should be from \eq{TM3}.

On the other hand, we can re-write \eq{TM4} in the form of the
 non-linear equation using \eq{TM5}: viz.

\be \label{TM7}
  \,\,\frac{\partial\,Z}{\partial \,Y}\,\,=\,\,-\,\Delta 
\left(Z  \,\,-\,\, Z^2\right) .
\ee

Comparing \eq{SOLTM} with \eq{ZF} one can see that
\be \label{TM8}
P_n\Lb Y \Rb\,\,=\,\,e^{\,-\,\Delta\, Y }\Lb 1\,-\, e^{\,-\,\Delta\,Y}\Rb^{n - 1} .
\ee

Since from \eq{TM8} it follows that the average  $n \,=\,N$ is equal
 to $N\,= \exp\Lb \Delta\,Y\Rb$ \eq{TM8} can be re-written in the
 form of \eq{I3}:
\be \label{NDIST}
P_n\Lb N\Rb\,\,=\,\,\frac{1}{N}\,\Big( 1\,-\,\frac{1}{N}\Big)^{n - 1}
\ee
  %%%%%%%%%%%%%%%%%%%%%%%%%%%%%%%%%%%%%%%%%%%%%%%%%%%%%%%%%%%%%%%
     \begin{figure}[ht]
     \begin{center}
     \includegraphics[width=  8cm,height=4.5cm]{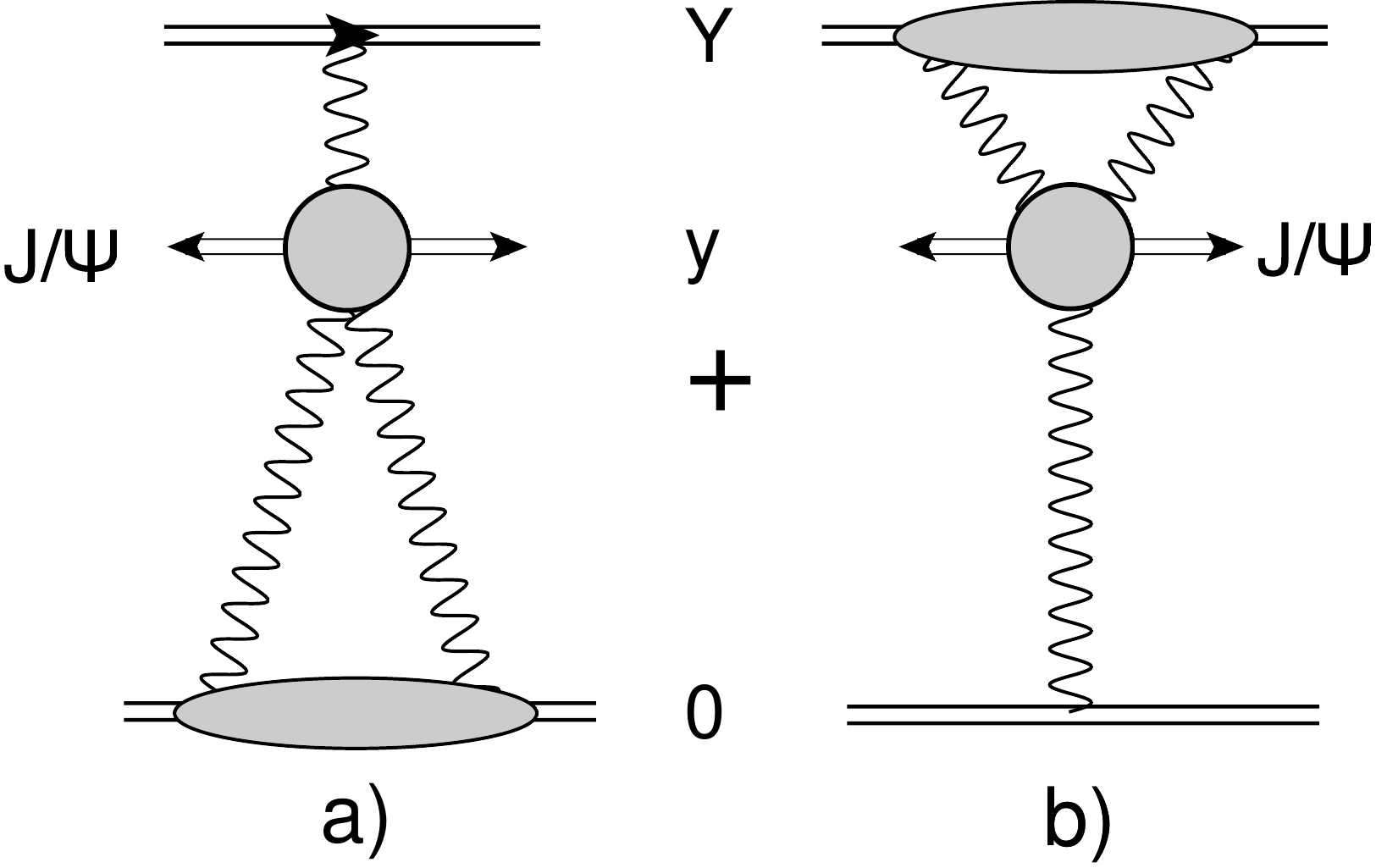} 
     \end{center}  
   \caption{ Mueller diagrams\cite{MUDIA}  for inclusive $J/\Psi$ production
 in the hadron-hadron collisions. The wavy lines denote the Pomeron Green's
 functions. }
\label{psipr}
   \end{figure}
%%%%%%%%%%%%%%%%%%%%%%%%%%%%%%%%%%%%%%%%%%%%%%%%%%%%%%% 

%%%%%%%%%%%%%%%%%%%%%%%%%%%%%%%%%%%%%%%%%%%
\subsection{Quarkonia production}
%%%%%%%%%%%%%%%%%%%%%%%%%%%%%%%%%%%%%%%%%%%%

As we have discussed in the introduction, we assume  the production of
 heavy quakonia stems from three gluon fusion (see \fig{3p}), and it is
 intimately related to the triple Pomeron interaction.   The Mueller
 diagrams for inclusive   $J/\Psi$ production are shown in \fig{psipr},
 where the wavy lines denote the Pomeron Green's function ($G_\pom$) which
 is equal to 
\be \label{POM}
G_\pom\Lb Y\Rb\,\,=\,\,e^{\Delta\,Y}
\ee

\begin{figure}[ht]
     \begin{center}
     \begin{tabular}{c c}
     \includegraphics[width=  0.5\textwidth]{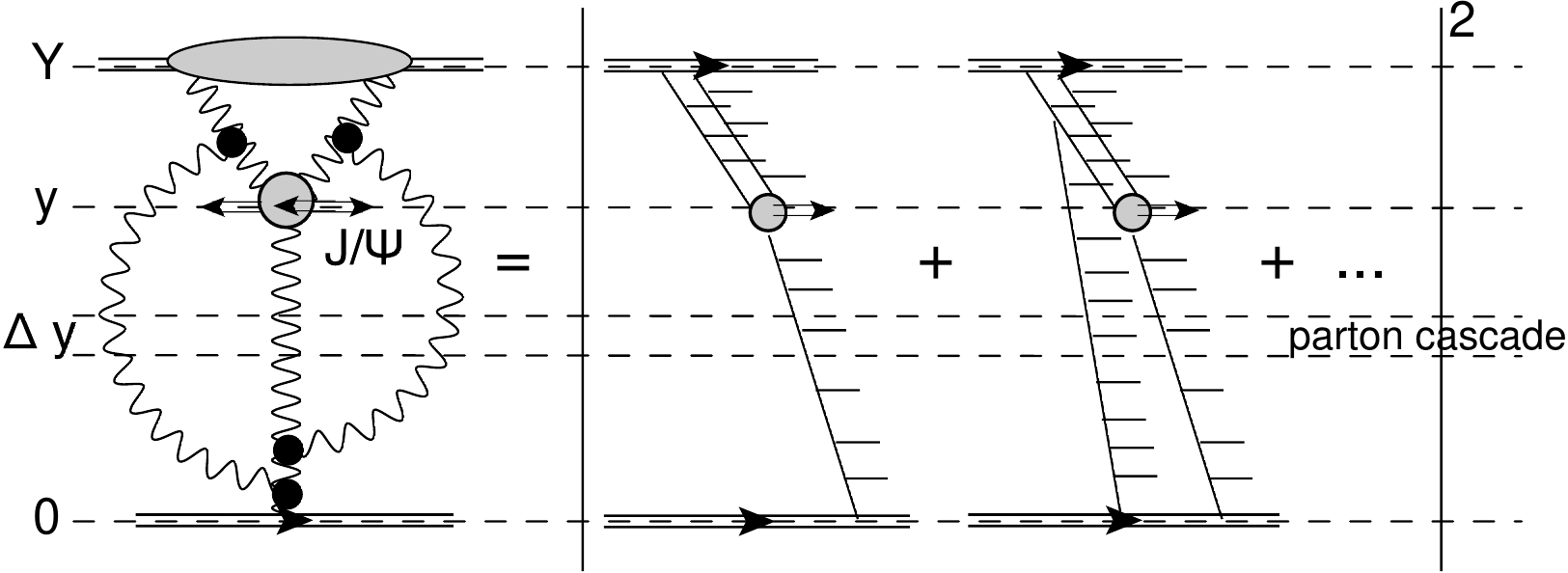} &
        \includegraphics[width=  0.5\textwidth]{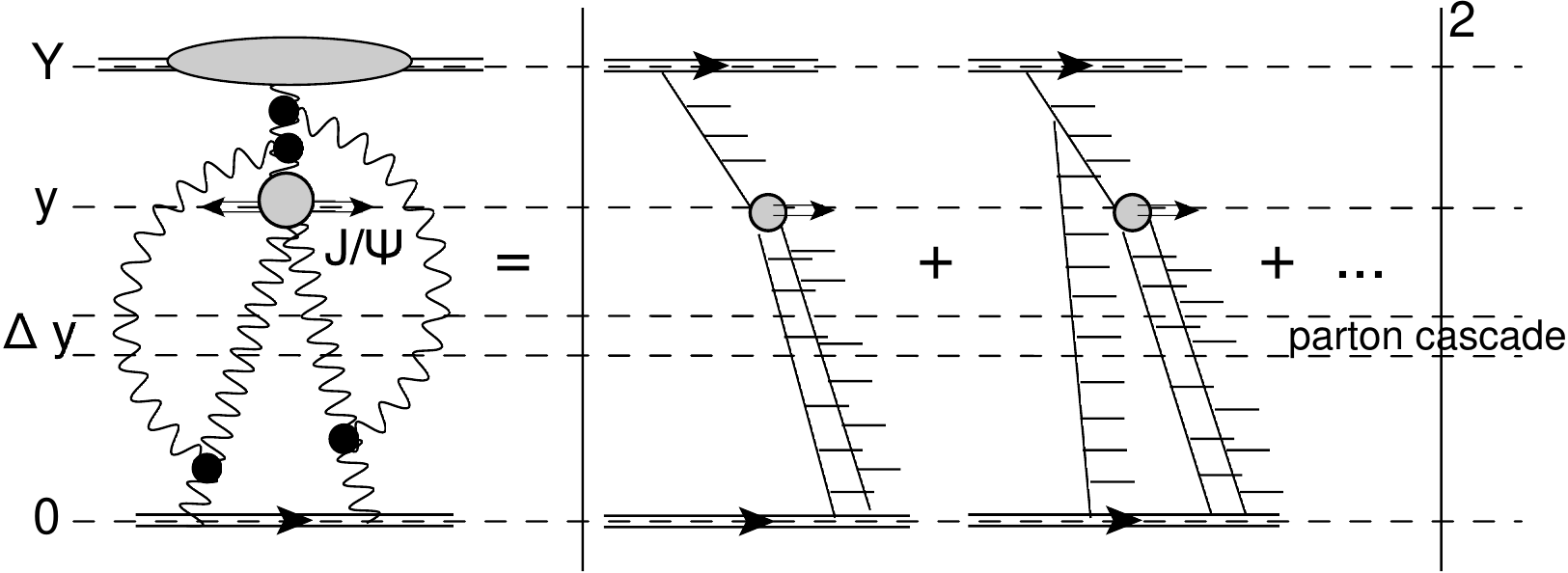} \\ 
           \fig{psiinmpsi}-a &  \fig{psiinmpsi}-b\\
         \end{tabular}
            \end{center}  
   \caption{ The example of Mueller diagrams\cite{MUDIA}  for inclusive
 $J/\Psi$ production
 in  hadron-hadron collisions with fixed multiplicity of produced hadrons
 and the parton cascade, which they describe.
 The exchange of Pomerons do not cancel each other due to AGK cutting
 rules\cite{AGK}, since we fixed the multiplicity in the final state. 
The wavy lines denote the Pomeron Green's functions. The black circles
 indicate the triple  Pomeron vertices. 
The solid lines correspond to partons. $\Delta y$ is the rapidity
 window in which the multiplicity of the soft hadron is measured. 
 This window is situated in central rapidity region with rapidity about $\h Y$.}
\label{psiinmpsi}
   \end{figure}

The MPSI approach for inclusive production with fixed multiplicity of
 produced hadrons is shown in \fig{psiinmpsi}

 Therefore, from this figure we see that  the structure of the parton
 cascade for the quarkonia production is quite different. In particular,
 for the part of the events whose weight is determined by the contribution
 of \fig{psipr}-b,
 the initial conditions for the parton cascade is not the ones of \eq{TM3} 
however, 
 they have the form:
 \be \label{TM9}
Z\Lb Y = 0,u\Rb \,\,=\,\,u^2;~~~~~~~~~Z\Lb Y, u=1\Rb\,\,=\,\,1.
\ee 

This  means, that for this cascade we need to find the arbitrary function
 $Z\Lb z\Rb$ from the following equation
\be \label{TM10}
Z\Lb z\Lb Y=0\Rb\Rb\,=\,u^2;  
\ee

The solution is
\be \label{SOLTM2}
Z\Lb Y, u\Rb\,\,=\,\,\frac{u^2\,e^{- 2\,\Delta \,Y}}{\Lb1 \,\,+\,\,u\Lb e^{ - \,\Delta\,Y}\,\,-\,\,1\Rb\Rb^2}\,\,=\,\,u^2\,e^{\,-\,2\,\Delta\, Y}\,\sum^\infty_{n=1}\Lb n \,-\,1\Rb\,u^n \Lb 1\,-\, e^{\,-\,\Delta\,Y}\Rb^{n -2}
\ee

    %%%%%%%%%%%%%%%%%%%%%%%%%%%%%%%%%%%%%%%%%%%%%%%%%%%%%%%%%%%%%%%
%%%%%%%%%%%%%%%%%%%%%%%%%%%%%%%%%%%%%%%%%%%%%%%%%%%%%%% 
   
      \eq{SOLTM2} leads to a different multiplicity distribution in
 comparison with \eq{NDIST}: viz.
      
      \be \label{N2DIS}
      P^{(2)}_n\,\,=\,\,\frac{1}{N^2} \Lb n\,\,-\,\,1\Rb \Lb 1 - \frac{1}{N}\Rb^{n - 2}
      \ee
      
       Finally, the  cross section for  quarkonia production  with
 given multiplicity ($n$)  of produced hadrons in the rapidity window
 $\Delta y$ (see \fig{psiinmpsi}), is
 equal  to 
      \be \label{NRDIST}
      \frac{d \sigma_n^{\rm J/\Psi}}{ d y} \,\,\,=\,\,   
 \frac{d \sigma^{\rm J/\Psi}_{\rm incl}}{ d y}\Lb
 \fig{psipr}-a\Rb \frac{n}{\langle n^{(1)}\rangle}
 \, \,  P^{(1)}_n\Lb N\Rb    \,\,\,+\,\,\, \frac{d
 \sigma^{\rm J/\Psi}_{\rm incl}}{ d y}\Lb \fig{psipr}-b\Rb
 \frac{n}{\langle n^{(2)}\rangle   }   \, \,  P^{(2)}_n\Lb N\Rb    \ee
  where $   \langle n^{(1)}\rangle $ and $ \langle n^{(2)}\rangle
 $ are average multiplicities of the produced hadrons in  the parton
 cascades  which    are shown in \fig{psiinmpsi}-a and in 
 \fig{psiinmpsi}-b, respectively.

In \eq{NRDIST} the first and the second terms correspond to  parton
 cascades  that    are shown in \fig{psiinmpsi}-a and in  \fig{psiinmpsi}-b,
 respectively. The appearance of the factor    $n/\langle n^{(i)}\rangle$ ,
 which is the number of parton ladders,  
 stems from the fact that $J/\Psi$ can be produced from every
  parton ladder (cut Pomeron) \cite{FEPA,MTVW,LSS}. It should be stressed
  that the number of ladders at rapidity $y$ from which the $J/\Psi$ is
 produced, is the same as the number of ladders from which the soft
 hadrons are produced in the MPSI approach\cite{MPSI}. It follows from
 the fact that integration over rapidities  ($y'$) of the triple Pomeron
 vertices in \fig{psiinmpsi} leads to $Y - y' \propto 1/\Delta$ for
 Pomerons in upper part of the diagram, and to $y' \propto 1/\Delta$
  in the lower part of the diagram. $\Delta$ denotes  the Pomeron 
intercept.

 Summing \eq{NRDIST} we obtain that 
 \be \label{NRDIST01}
 \sum_n    \frac{d \sigma_n^{\rm J/\Psi}}{ d y} \,\,=\,\,
  \frac{d \sigma^{\rm J/\Psi}_{\rm incl}}{ d y}\Lb \fig{psipr}-a\Rb\,\,+\,\,   \frac{d \sigma^{\rm J/\Psi}_{\rm incl}}{ d y}\Lb \fig{psipr}-b\Rb\,\,=\,\,\frac{d \sigma^{\rm J/\Psi}_{\rm incl}}{ d y}
 \ee
 which coincide with  the Mueller diagram approach, shown in \fig{psipr}.
 Introducing $\kappa\,\, =  \frac{d \sigma^{\rm J/\Psi}_{\rm incl}}{ d y}\Lb
 \fig{psipr}-b\Rb \Bigg{/}  \frac{d \sigma^{\rm J/\Psi}_{\rm incl}}{ d y}\Lb
 \fig{psipr}-a\Rb  $ we re-write \eq{NRDIST} in the following form:
     \be \label{NRDIST02}
      \frac{\frac{d \sigma_n^{\rm J/\Psi}}{ d y}}{\frac{d
 \sigma^{\rm J/\Psi}_{\rm incl}}{ d y}} \,\,\,=\,\, 
 \frac{1}{1\,\,+\,\,\kappa}\Bigg( \frac{n}{\langle n^{(1)}\rangle}
 \, \,  P^{(1)}_n\Lb N\Rb    \,\,\,+\,\,\, \kappa \frac{n}{\langle
 n^{(2)}\rangle   }   \, \,  P^{(2)}_n\Lb N\Rb\Bigg)    \ee

      where  $\kappa$  is equal to \cite{LESI}
      \be \label{KAPPA}
      \kappa\,\,=\,\,e^{ 2\,\Delta\Lb \h Y \,\,-\,\,y\Rb}
      \ee
      In \eq{NRDIST} $P^{(1)}_n\Lb N\Rb\,\,\equiv\,\,P_n\Lb N\Rb$ of
 \eq{NDIST}.      
      
       The cross section of produced gluons (hadrons) is proportional to
       \be \label{TM11}
       \frac{d \sigma^{\rm prod. gl.}_n}{ d y} \,\,=\,\, \, \,   \frac{d \sigma^{\rm prod. gl.}_{\rm incl}}{ d y}  P^{(1)}_n\Lb N\Rb  
         \ee
         
\begin{figure}[ht]
     \begin{center}
     \begin{tabular}{c c c}
     \includegraphics[width=  0.45\textwidth]{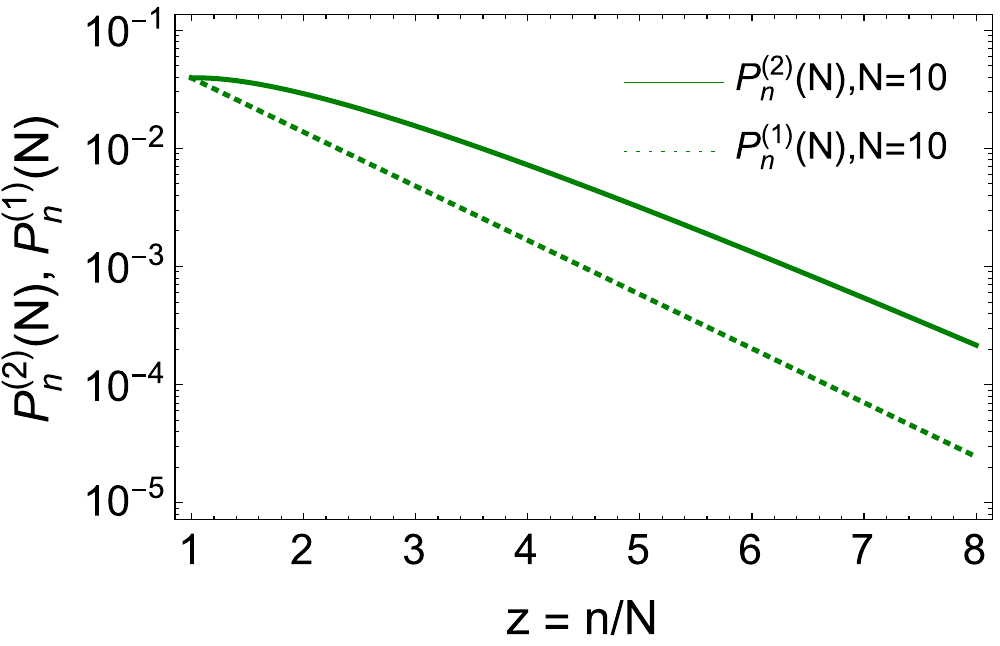} &~~~~&  \includegraphics[width= 0.42\textwidth]{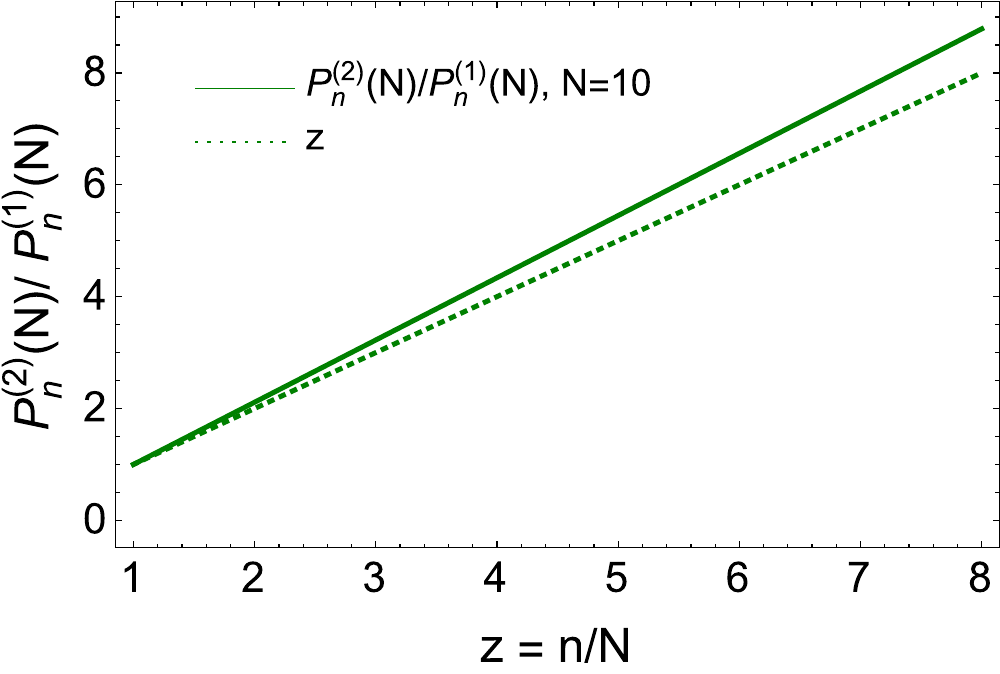}  \\
     \fig{p2p1}-a & &\fig{p2p1}-b\\
     \end{tabular}
     \end{center}  
   \caption{\fig{p2p1}-a:  $P^{(2)}_n\Lb N\Rb$ and $P^{(1)}_n\Lb N\Rb$
 versus $z = \frac{n}{N}$. $N$ is taken to be equal 10.   \fig{p2p1}-b:
 The ratio of $P^{(2)}_n\Lb N\Rb/P^{(1)}_n\Lb N\Rb$  for $N=10$ and for
 large $N$ (see \eq{TM13}}
   \label{p2p1}
   \end{figure}

From \fig{p2p1}, one can see that $P^{(2)}_n\Lb N\Rb $ and
 $P^{(1)}_n\Lb N\Rb$ have different dependance on $z = n/N$.
 The average number of gluons is chosen to be the  mean multiplicity
 of hadrons in the rapidity window $ |\eta|\,\leq\, 0.9$ measured at
 W=13TeV.

   %%%%%%%%%%%%%%%%%%%%%%%%%%%%%%%%%%%%%%%%%%%%%%%%%%%%%%%%%%%%%%%
%%%%%%%%%%%%%%%%%%%%%%%%%%%%%%%%%%%%%%%%%%%%%%%%%%%%%%% 
       Therefore,     $   \frac{d \sigma^{\rm J/\Psi}}{ d y}  $ is not
 proportional to  $  \frac{d \sigma^{\rm prod.\, gl.i}}{ d y}$, but
 shows non-linear dependance,  which we will discuss below.    It
 should be stressed that such dependance stems from triple Pomeron
 mechanism of       quarkonia production,  as  noted
 in Refs.\cite{LESI,LSS}. At large $N$ we have
       \be \label{TM12}
      P^{(2)}_n\,\,\xrightarrow{N\,\gg\,1}\,\,\frac{1}{N} z\,e^{-z};~~~~~~~~ P^{(1)}_n\,\,\xrightarrow{N\,\gg\,1}\,\,\frac{1}{N} \,e^{-z};      \ee
   leading to
        \be \label{TM13}
     \frac{ P^{(2)}_n}{P^{(1)}_n} \,\,\xrightarrow{N\,\gg\,1}\,\,z;  
        \ee

      It is worth mentioning that the average multiplicity of $P^{(2)}$
 distribution is equal to
      \be \label{TM14}
      < n^{(2)} >\,\,=\,\,\sum^\infty_{n=2}\,n\, P^{(2)}_n\Lb N\Rb\,\,=\,\,2\,N
      \ee
      Hence, for the multiplicity distribution $P^{(2)}$ the average
 number of accompanying gluons(hadrons) is twice larger than in the
 distribution $P^{(1)}$ .      \eq{TM14} means that the ratio
 $\frac{n^{(2)}}{ < n^{(2)} > }\,\,=\,\,\frac{n}{ 2\,N}$.

   %  %%%%%%%%%%%%%%%%%%%%%%%%%%%%%%%%%%%%%%%%%%%%%%%%%%
\section{Structure of QCD parton cascade}
  %%%%%%%%%%%%%%%%%%%%%%%%%%%%%%%%%%%%%%%%%%%   

  %%%%%%%%%%%%%%%%%%%%%%%%%%%%%%%%%%%%%%%%%%%%%%%%%%%%%
-    
  \subsection{Multiplicity distribution}         
  
  %%%%%%%%%%%%%%%%%%%%%%%%%%%%%%%%%%%%%%%%%%%%%%%%%%%
  
   In QCD, to find the multiplicity distribution for hadron-hadron 
scattering in QCD using MPSI approach\cite{MPSI}, we 
  need to evaluate (see \eq{Z})
 \be \label{QCD1}
\tilde{P}_n\Lb Y, r\Rb\,\,=\,\,
\int P_n\Lb Y,\vec{r},\vec{b};\{\vec{r}_i\,\vec{b}_i\}\Rb \prod^{n}_{i=1} \,d^2 r_i\,d^2 b_i
\ee  
  
  However, in the MPSI approach it  is more natural  to introduce moments
 (see \eq{N3}):
\be \label{QCD2}
M^p_n\Lb Y,  r\Rb\,\,=\,\,\int_{r}  \prod^n_{i=1} d^2 r_i  d^2 b \,\rho^p_n\Lb  Y,  \{ r_i\},b\Rb\,\,=\,\,\int_{r}  \prod^n_{i=1}\frac{ d^2 r_i }{r^2_i }  d^2 b\, \bar{\rho}^p_n\Lb Y, \{r_i\},b\Rb
\ee
  The integration over $r_i$ depends on the size of the initial dipole,
 which generates the cascade.
  In DIS the natural integration stems from $r_i \,>\,r$.  
  
    %%%%%%%%%%%%%%%%%%%%%%%%%%%%%%%%%%%%%%%%%%%%%%%%%%%%%
  
  \subsubsection{Several first iterations. }         
  
  %%%%%%%%%%%%%%%%%%%%%%%%%%%%%%%%%%%%%%%%%%%%%%%%%%% 
  
  We start from the first several iteration of \eq{N3}, which can
 be re-written for $\,\bar{\rho}^p_n(r_1, b_1\,\ldots\,,r_n, b_n)$
 in the form
  
   \bea \label{QCD3}
\frac{\partial \,\bar{\rho}^p_n(r_1, b_1\,\ldots\,,r_n, b_n)}{ 
\bar{\alpha}_s\,\partial\,Y}\,\,&=&\,\
-\,\sum_{i=1}^n
 \,\,\omega(r_i)\,\,\bar{\rho}^p_n(r_1, b_1\,\ldots\,,r_n, b_n)\,\,+\,\,2\,\sum_{i=1}^n\,
\int\,\frac{d^2\,r'}{2\,\pi}\,
\frac{1}{(\vec{r}_i\,-\,\vec{r}')^2}\,
\bar{\rho}^p_n(\ldots\,r', b_i-r'/2\dots)\nn\\
 & & 
\,
+\,\sum_{i=1}^{n-1}\,
\bar{\rho}^p_{n-1}(\ldots\,(\vec{r}_i\,+\,\vec{r}_n), b_{in}\dots).
\eea  
  For the first iteration $\bar{\rho}_1\Lb Y; r_1,b_1\Rb$, we obtain the
 BFKL equation:
  \be \label{QCD4}
\frac{\partial \,\bar{\rho}^p_1(Y; r_1, b)}{ 
\bas\,\partial\,Y}\,\,=\,\,-\,\,\omega_G\Lb r_1\Rb\bar{\rho}^p_1(Y; r_1, b) \,\,+\,2\,\int\,\frac{d^2\,r'}{2\,\pi}\,
\frac{1}{(\vec{r}_1\,-\,\vec{r}')^2}\,
\bar{\rho}^p_1\Lb Y, r',b\Rb\,\,=\,\,\int d^2 \,r' K\Lb r_1,r'\Rb \bar{\rho}^p_1\Lb Y, r',b\Rb\ee  
  with the solution
    \be \label{QCD5}  
  \bar{\rho}^p_1(Y; r_1, b)\,\,=\,\,\int^{\epsilon\,+\,i\,\infty}_{\epsilon \,-\,i\,\infty} \frac{d \omega}{ 2\,\pi\,i}  
  \int^{\epsilon\,+\,i\,\infty}_{\epsilon \,-\,i\,\infty} \frac{d \gamma}{ 2\,\pi\,i}  \,e^{ \omega\,Y\,\,+\,\,\gamma\,\xi_1}\frac{1}{\omega\,-\,\bas \chi\Lb \gamma\Rb} \tilde{\rho}^p_{in,1}(\gamma, b)  \ee
  where $\xi_1\,\,=\,\,\ln\Lb r^2_1\Lambda_{\rm QCD}^2\Rb$ and 
\bea \label{BFKLKER}
\omega\Lb  \gamma\Rb\,\,&=&\,\,\bas\,\chi\Lb \gamma \Rb\,\,\,=\,\,\,\bas \Lb 2 \psi\Lb 1\Rb \,-\,\psi\Lb \gamma\Rb\,-\,\psi\Lb 1 - \gamma\Rb\Rb\\
\mbox{diffusion approximation}&\,\xrightarrow{\gamma \to \h}& \,\omega_0\,\,+\,\,D\,\Lb \gamma - \h\Rb^2  \,\,+\,\,{\cal O}\Lb (\gamma - \h)^3\Rb\,\,\nn\\
&=&\,\,\bas 4 \ln 2  \,\,+\,\,\bas 14 \zeta\Lb 3\Rb \Lb \gamma - \h\Rb^2  \,\,+\,\,{\cal O}\Lb (\gamma - \h)^3\Rb\nn
 \eea

where $\psi(z)$ is Euler gamma function (see \cite{RY} formula{ \bf 8.36}). $\tilde{\rho}^p_{in,1}(\gamma, b) $ has to be found from the initial conditions. From \eq{QCD5} we can obtain $  M^p_1\Lb Y,r,b\Rb$ (see \eq{QCD2}) which has the following form:
  \be \label{QCD6}  
M^p_1\Lb Y,r\Rb\,\,=\,\,\int^{\epsilon\,+\,i\,\infty}_{\epsilon \,-\,i\,\infty} \frac{d \omega}{ 2\,\pi\,i}  
  \int^{\epsilon\,+\,i\,\infty}_{\epsilon \,-\,i\,\infty} \frac{d \gamma}{ 2\,\pi\,i}  \,e^{ \omega\,Y\,\,+\,\,\gamma\,\xi}\frac{1}{\omega\,-\,\bas \chi\Lb \gamma\Rb} M^p_{in,1}(\gamma)  \ee  
  which satisfies the following equation:
  \be \label{QCD60}
\frac{ \partial\, M^p_1\Lb Y, r\Rb}{\partial\,\bas\,Y}\,\,=\,\,\,\,\,\int d^2 \,r' K\Lb r,r'\Rb M^p_1\Lb Y, r'\Rb
\,\,\xrightarrow{r'\,\gg\,r}\,\,\,\int_r\frac{d \,r'^2}{r'^2} \,\,M^p_1\Lb Y, r'\Rb
\ee  
  
  The equation for the next iteration: $\bar{\rho}_2$,  takes the form:
  Equation for $\rho^p_2$ can be re-written in the following form for
 $\bar{\rho}^p_2$:
 \begin{subequations}
 \bea 
&&\hspace{-1cm}\frac{\partial \,\bar{\rho}^p_2(Y; r_1, r_2, b)}{ 
\bas\,\partial\,Y}\,\,=\,\,\int d^2\,r'\,K\Lb r_1, r'\Rb 
\bar{\rho}^p_2\Lb Y, r',b, r_2,b\Rb\,+\,\,\,\int  d^2\,r'\,K\Lb r_2, r'\Rb
\bar{\rho}^p_2\Lb Y, r_1, r', b\Rb\,\,+\,\,\bar{\rho}^p_{1}\Lb Y; \vec{r}_1\,+\,\vec{r}_2, b\Rb\label{QCD70}\\
&&~~~~~~~~~~~\xrightarrow{r'\,\gg\,r_i} \int_{r_1} \frac{d r'^2}{r'^2} \bar{\rho}^p_2\Lb Y, r',b, r_2,b\Rb\,\,+\,\,
\int_{r_2} \frac{d r'^2}{r'^2} \bar{\rho}^p_2\Lb Y, r_1,b, r',b\Rb\,\,+\,\,\bar{\rho}^p_{1}\Lb Y; \vec{r}_1\,+\,\vec{r}_2, b\Rb\label{QCD71}
\eea
 \end{subequations}

For simplicity we re-write \eq{QCD70}  in the  log approximation
 following Ref.\cite{GOLEMULT} (see \eq{QCD71}).
Rewriting \eq{QCD71} for $M^p_2$ we obtain:
\be \label{QCD8}
\frac{\partial \,M^p_2(Y; r)}{ 
\bas\,\partial\,Y}\,\,=\,\, \int_r \frac{d r'^2}{r'^2} \,\,\Bigg\{ 2\,M^p_2\Lb Y, r' \Rb \,\,+\,\,M^p_1\Lb Y, r'\Rb\Bigg\}\,\,=\,\, 
2\,\int_r \frac{d r'^2}{r'^2} \,\,\,M^p_2\Lb Y, r' \Rb\,\,+\,\,\frac{\partial}{ \partial\,\bas\,Y} M^p_1\Lb Y, r \Rb
\ee
In the last term of \eq{QCD8} we used \eq{QCD60}. The solution to
 \eq{QCD8} takes the form:

  \be \label{QCD9}
 M^p_2\Lb Y, r\Rb\,\,\,=\,\,\,\int^{\epsilon\,+\,i\,\infty}_{\epsilon \,-\,i\,\infty} \frac{d \omega}{ 2\,\pi\,i}  
  \int^{\epsilon\,+\,i\,\infty}_{\epsilon \,-\,i\,\infty} \frac{d \gamma}{ 2\,\pi\,i}    \,e^{ \omega\,Y\,\,+\,\,\gamma\,\xi}\frac{\bas \chi\Lb \gamma\Rb }{\Lb \omega\,\,-\,\,2\,\bas \chi\Lb \gamma\Rb\Rb\,\Lb \omega\,\,-\,\,\bas \chi\Lb \gamma\Rb\Rb} 
   \ee 
with $\chi\Lb \gamma\Rb= 1/\gamma$. One can check that
 $ M^p_2\Lb Y, r\Rb \,\xrightarrow{Y \,\to\,0}\,\,0$, which
 is the correct initial condition for one dipole of size $r$
  at $Y=0$ ,  which generates the parton cascade.
Actually, \eq{QCD8} describes $M^p_2\Lb Y, r\Rb$  for the full
 BFKL kernel.  Indeed, considering \eq{QCD70} we can integrate
 this equation over $r_1$ and $r_2$, to obtain the equation for
 $M^p_2$. The last term has the following form
\be \label{QCD10}
\int_r \frac{d^2 r_1}{2\,\pi} \frac{1}{r^2_1} \frac{d^2 r_2}{2\,\pi}\int d^2 b  \frac{1}{r^2_2} \bar{\rho}(Y,r_{12}, b)\,=\,
\int_r \frac{d^2 r_1}{2\,\pi} \frac{1}{r^2_1} \frac{d^2 r_2}{2\,\pi}\int d^2 b \, \frac{1}{\Lb \vec{r}_1 \,-\,\vec{r}_{12}\Rb^2} \,\bar{\rho}(Y,r_{12}, b)
\ee
 Note that $r^2_2\,=\,\Lb \vec{r}_1 \,-\,\vec{r}_{12}\Rb^2$
  can never approach zero, since $r_2> r$. Removing this restriction, we
 can re-write
\be \label{QCD11}
\int_r  \frac{d^2 r_2}{2\,\pi}\, \frac{1}{\Lb \vec{r}_1 \,-\,\vec{r}_{12}\Rb^2} \,\bar{\rho}(Y,r_{12}, b)\,\,=\,\,
\int_0  \frac{d^2 r_2}{2\,\pi}\, \frac{1}{\Lb \vec{r}_1 \,-\,\vec{r}_{12}\Rb^2} \,\bar{\rho}(Y,r_{12}, b)\,\,-\,\,\ln r^2_1\,
\bar{\rho}(Y,r_{1}, b)
\ee
The reggeization term in \eq{QCD11} describes the contribution of
 $r_2\,\to\,0$. Plugging \eq{QCD11} in the last term of \eq{QCD70}
 integrated over $r_1$ and $r_2$, one can see that we reproduce \eq{QCD8}
 for the full BFKL kernel. Hence \eq{QCD9} is the solution with
 $\chi\Lb\gamma\Rb$ which is given by \eq{BFKLKER}.
    %%%%%%%%%%%%%%%%%%%%%%%%%%%%%%%%%%%%%%%%%%%%%%%%%%%%%
  
  \subsubsection{General solution  }         
  
  %%%%%%%%%%%%%%%%%%%%%%%%%%%%%%%%%%%%%%%%%%%%%%%%%%% 

The equation for $M^p_n\Lb Y, r\Rb$ has the following general form:

\bea \label{QCD12}
&&\frac{\partial \,M^p_n(Y; r)}{ 
\bas\,\partial\,Y}\,\,=\\
&&\,\, \int  d^2\,r'\,K\Lb r, r'\Rb \,\,\Bigg\{ n\,M^p_2\Lb Y, r' \Rb \,\,+\,\,(n \,-\,1)M^p_1\Lb Y, r'\Rb\Bigg\}\,\,=\,\, 
n\,\int  d^2\,r'\,K\Lb r, r'\Rb  \,\,\,M^p_2\Lb Y, r' \Rb\,\,+\,\,\Lb n\,-\,1\Rb\frac{\partial}{ \partial\,\bas\,Y} M^p_1\Lb Y, r \Rb\nn
\eea
The solution to this equation\cite{GOLEMULT}, which gives $M^p_1\Lb Y=0,
 r\Rb \,=\,1$, but all other $M^p_n$ with $n\,\geq\, 2$=0, are equal to

\be \label{QCDSOL1}
M^p_n\Lb Y, r\Rb\,\,=\,\,\int^{\epsilon\,+\,i\,\infty}_{\epsilon\,-\,i\,\infty}\frac{d \gamma}{2\,\pi\,i}e^{  \bas \chi\Lb \gamma\Rb\,Y}\,\Bigg( e^{  \bas \chi\Lb \gamma\Rb\,Y}\,\,-\,\,1\Bigg)^{n - 1}
\ee
which leads to the multiplicity distribution, which takes the form (see
 \eq{QCD1} and Ref.\cite{GOLEMULT}):
\be \label{QCDSOL2}
\tilde{P}_n\Lb Y, r \Rb\,\,\,=\,\,\int^{\epsilon\,+\,i\,\infty}_{\epsilon
\,-\,i\,\infty}\frac{d \gamma}{2\,\pi\,i}\,
e^{-\,  \bas \chi\Lb \gamma\Rb\,Y}\,\Bigg( 1\,\,-\,\,e^{ -\, \bas \chi\Lb
 \gamma\Rb\,Y}\Bigg)^{n - 1}
\ee

For $N\,\,=\,\,e^{  \bas \chi\Lb \gamma\Rb\,Y} \,\,\gg\,\,1$ we have

 \be \label{QCDSOL3}
\tilde{P}_n\Lb Y, r \Rb\,\,\,=\,\,\int^{\epsilon\,+\,i\,\infty}_{\epsilon\,-\,i\,\infty}\frac{d \gamma}{2\,\pi\,i}\,\exp\Lb - z(\gamma)\,\,+\,\,\gamma\,\xi\Rb~~~~\mbox{where}~~~~z\,\,=\,\,\frac{n}{e^{  \bas \chi\Lb \gamma\Rb\,Y} }\ee
Taking the integral over $\gamma$ using the method of steepest descent,
 using the diffusion approximation for the BFKL kernel (see \eq{BFKLKER}).
 The equation for $\gamma_{\rm SP}$ has the following form:
\be \label{SP}
2\,D\,Y z\Lb \h\Rb \Lb \gamma_{\rm SP} \,-\,\h\Rb\,+\,\xi\,=0~~~~\mbox{with}~~~\Lb \gamma_{\rm SP} \,-\,\h\Rb\,\,=\,\,-\,\frac{\xi}{2\,D\,Y\,z\Lb \h\Rb}
\ee
and the integral over $\gamma$  is
\be \label{QCDSOL30}
\tilde{P}_n\Lb Y, r \Rb\,\,\,=\,\,\, \sqrt{\frac{\pi}{2\,D\,z\Lb \h\Rb\,Y}}\, \frac{1}{ N\Lb Y\Rb}\,e^{ - z\Lb \h\Rb}\,
~~~\mbox{with} ~~~~z\,=\,\frac{n}{N\Lb Y\Rb}~~~~\mbox{and}~~~N\Lb Y\Rb\,\,=\,\,e^{\omega_0\,Y} 
\ee
considering $ \xi^2\Big{/}2\,D\,z\Lb\h\Rb Y \,\ll\,1$. After normalization,
 we obtain that
\be \label{QCDSOL4}
\frac{< n  > \,\sigma_n}{\sigma_{\rm in}}\,\,\,=\Psi^{\rm KNO} \Lb z\Rb\,\,\,=\,\,\, \sqrt{\frac{1}{\pi \,z}}\, \,e^{ - z\Lb \h\Rb}\,
~~~\mbox{with} ~~~~z\,=\,\frac{n}{N\Lb Y\Rb}~~~~\mbox{and}~~~N\Lb Y\Rb\,\,=\,\,e^{\omega_0\,Y} 
\ee
where $\Psi^{\rm KNO}$ denotes the KNO function (see Ref.\cite{KNO}).

It is worthwhile mentioning that the multiplicity distribution of
 \eq{QCDSOL4} is different from \eq{NDIST} and 
\be \label{QCDR}
R\,=\,\frac{P_n^{\rm QCD}\Lb \eq{QCDSOL4}\Rb}{P_n\Lb \eq{NDIST}\Rb}\,\,=\,\,\sqrt{\frac{1}{\pi \,z}}
\ee

\begin{figure}[ht]
     \begin{center}
     \includegraphics[width=  0.5\textwidth]{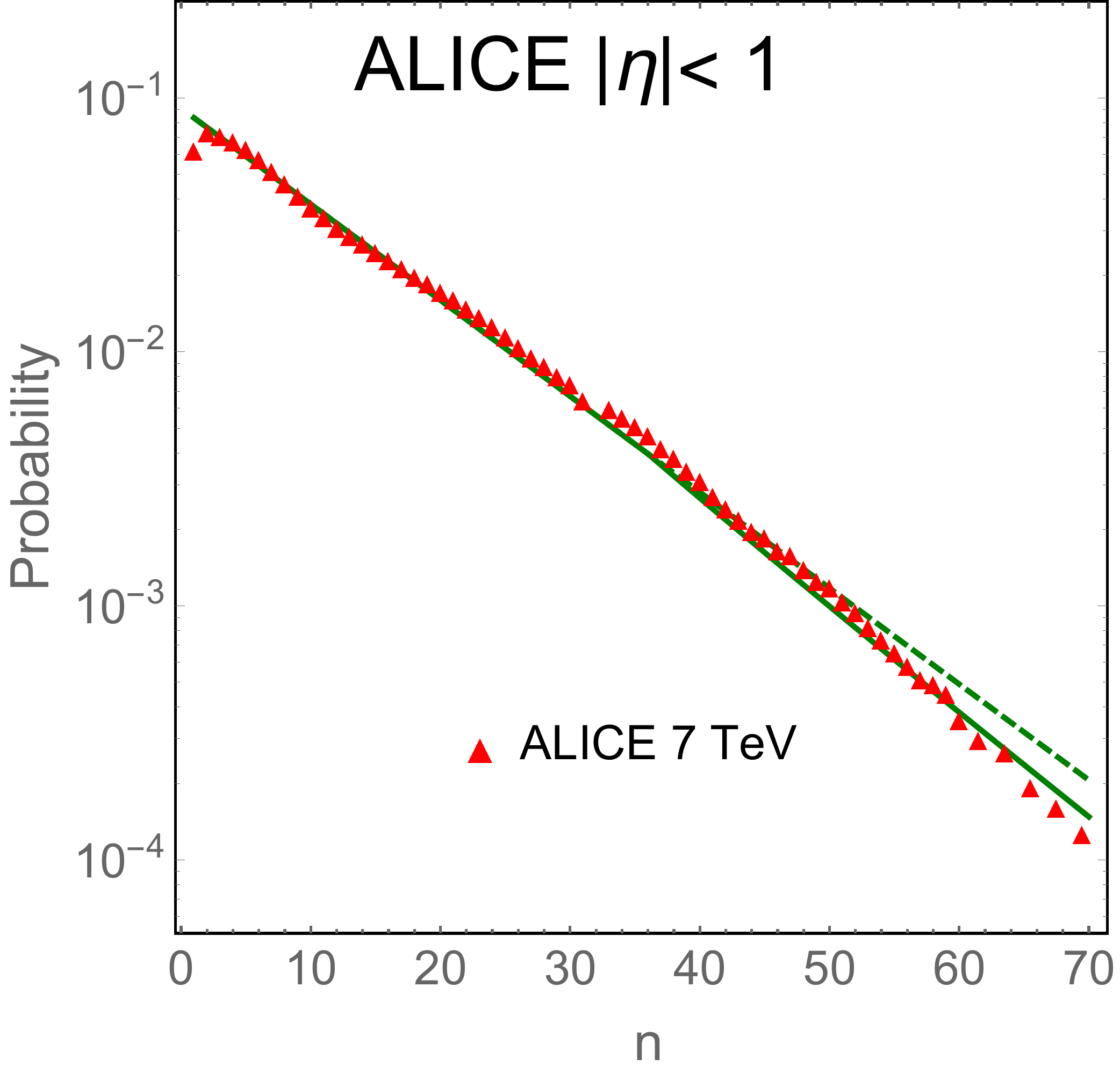} 
     \end{center}  
   \caption{Multiplicity distribution of the charged hadrons in the
 central rapidity region. The solid line  is the distribution
 of \eq{QCDSOL30}, while the dotted curve corresponds to \eq{NDIST}.
 The data  and the value of $N=12$ are taken from Refs.\cite{ALICEMULT}.  }
\label{mult}
   \end{figure}

In \fig{mult} we compare the ALICE data\cite{ALICEMULT} on multiplicity
 distribution with \eq{QCDSOL30} and with \eq{NDIST}. On can see that
 the agreement is good, and the difference between the above equations
 can be seen at large $n$. In describing the experimental data we use
 \eq{QCDSOL30}, which is derived at large $z$, for $z\,\geq\,3$.
It worthwhile mentioning that  the data of CMS \cite{CMSMULT} we have
 discussed in our paper\cite{GOLEMULT}.
    %%%%%%%%%%%%%%%%%%%%%%%%%%%%%%%%%%%%%%%%%%%%%%%%%%%%%%%%%%%%%%%
%%%%%%%%%%%%%%%%%%%%%%%%%%%%%%%%%%%%%%%%%%%%%%%%%%%%%%% 
  
  %%%%%%%%%%%%%%%%%%%%%%%%%%%%%%%%%%%%%%%%%%%%%%%%%%%%%
  \begin{boldmath}
  \subsection{ $P^{(2)}_n$ distribution for quarkonia production}   
  \end{boldmath}      
  
  %%%%%%%%%%%%%%%%%%%%%%%%%%%%%%%%%%%%%%%%%%%%%%%%%%%

The $P^{(1)}_n $ distribution , which we have discussed in the previous
 section, can be derived, using the double  Laplace transform
 representation, both for $P^{(1)}_n\Lb Y; r\Rb $  and for
 $M^{(1)}_n\Lb Y; r\Rb $:
\be \label{OMREP}
M^{(1)}_n\Lb Y; r\Rb\,\,\,=\,\,\,\int^{\epsilon\,+\,i \infty}_{\epsilon\,-\,i\,\infty}\frac{d \omega}{2\,\pi\,i}\int^{\epsilon\,+\,i \infty}_{\epsilon\,-\,i\,\infty}\frac{d \gamma}{2\,\pi\,i}\,e^{\omega\,Y\,\,+\,\,\gamma\,\xi}\,\, m^{(1)}_n\Lb \omega, \gamma\Rb
\ee
where $\xi\,=\,\ln\Lb r^2 \Lambda^2_{\rm QCD}\Rb$.

\eq{QCD12}  in the $\omega$-representation, has the following form:
\be \label{QCDOM1}
\omega\,m^{(1)}_n\Lb \omega, \gamma\Rb\,\,\,=\,\,n\,\bas \chi\Lb \gamma\Rb \,m^{(1)}_n\Lb \omega, \gamma\Rb
\,\,+\,\,\Lb n - 1 \Rb\,\bas \chi\Lb \gamma\Rb \,\,m^{(1)}_{n-1}\Lb \omega, \gamma\Rb
\ee
with the solution:

\be \label{QCDOM2}
m^{(1)}_n\Lb \omega, \gamma\Rb\,\,\,=\,\,\Lb n\, -\,1\Rb! \prod^{n}_{m=1}    \frac{   \bas \chi\Lb \gamma\Rb}{  \omega\,\,-\,\,m\,\bas\chi\Lb \gamma\Rb}
\ee

This solution gives $M^p_1\Lb Y=0,r\Rb\,\,\neq\,\,0$, while
 $M^p_n\Lb Y=0,r\Rb\,\,=\,\,0$ for $\,n\,\geq\,2$. The inverse
 Laplace transform leads to \eq{QCDSOL1} for $M^p_n\Lb Y, r\Rb$ and
 \eq{QCDSOL2}\,for $P^{(1)}_n\Lb Y, r\Rb$.

To find  $P^{(2)}_n$ distribution we need to take into account that at $Y=0$:

\be \label{QCDOM3}   
M^p_2\Lb Y  =  0, r\Rb\,\,=\,\,1;~~~~~~~~M^p_1\Lb Y  =  0, r\Rb\,\,=\,\,0;~
~~~~~~M^p_n\Lb Y  =  0, r\Rb\,\,=\,\,0~~\mbox{for}~~n\,\,\geq\,\,3
\ee
One can see that the following: $m^{(1)}_1\Lb \omega, \gamma\Rb$ satisfies
 thess conditions:

\be \label{QCDOM4}
m^{(2)}_n\Lb \omega, \gamma\Rb\,\,\,=\,\,\Lb n\, -\,1\Rb! \prod^{n}_{m=2}    \frac{   \bas \chi\Lb \gamma\Rb}{  \omega\,\,-\,\,m\,\bas\chi\Lb \gamma\Rb}
\ee

The inverse Laplace transform with respect to $\omega$ leads to
\be \label{QCDOM5}
M^{(1)}_n\Lb Y; r\Rb\,\,\,=\,\,\,\int^{\epsilon\,+\,i \infty}_{\epsilon\,-\,i\,\infty}\frac{d \gamma}{2\,\pi\,i}\,e^{\gamma\,\xi}\,\,\frac{1}{\gamma}\,\Lb n\,-\,1\Rb\, e^{2\,\bas\,\chi\Lb \gamma\Rb\,Y}\,\Big(e^{2\,\bas\,\chi\Lb \gamma\Rb\,Y}\,\,-\,\,1\Big)^{n\,-\,2}
\ee
which gives the initial conditions of \eq{QCDOM3}.

For  $P^{(2)}_n$ we obtain
\be \label{QCDOM6}
P^{(2)}_n\Lb Y; r\Rb\,\,\,=\,\,\,\int^{\epsilon\,+\,i \infty}_{\epsilon\,-\,i\,\infty}\frac{d \gamma}{2\,\pi\,i}\,e^{\gamma\,\xi}\,\,\frac{1}{\gamma}\,\Lb n\,-\,1\Rb\, e^{-\,2\,\bas\,\chi\Lb \gamma\Rb\,Y}\,\Big(1\,\,-\,\,e^{-2\,\bas\,\chi\Lb \gamma\Rb\,Y}\Big)^{n\,-\,2}
\ee
with $P^{(2)}_2\Lb Y=0,r\Rb\,\,=\,\,1$ and
 $P^{(2)}_n\Lb Y=0,r\Rb\,\,=\,\,0$ for $n\,\neq\,2$ at $Y=0$.

Repeating the same estimates as in \eq{QCDSOL3}, we obtain
 the KNO function,
\be \label{QCDOM7}
\Psi^{KNO}\Lb z\Rb\,\,=\,\,2\,\sqrt{ \frac{z}{\pi}}\,e^{-\,z}
\ee
with the normalization $\int d z\,\Psi^{KNO}\Lb z\Rb\,\,=\,\,1$.

Note  that the ratio 
 $\frac{ P^{(2)}_n\Lb Y, r\Rb}{P^{(1)}_n\Lb Y , r\Rb} \,\,=\,\,z$ for 
large $z$,  as in \eq{TM13}.

~

~

    %  %%%%%%%%%%%%%%%%%%%%%%%%%%%%%%%%%%%%%%%%%%%%%%%%%%
\section{Comparison with experimental data}
  %%%%%%%%%%%%%%%%%%%%%%%%%%%%%%%%%%%%%%%%%%~

In both ALICE\cite{ALICE0,ALICE1,ALICE01,ALICE2,ALICE3} and STAR
 \cite{STAR1,STAR2} experiments the following ratio is measured:
\be \label{COM1}
\frac{n_{J/\Psi}}{\langle n_{J/\Psi}\rangle } \,\,=\,\,F\Lb \frac{n}{N}\Rb
\ee
where $N\,=\langle n \rangle$ is the average number of charged hadrons
 in the fixed rapidity window, and $\langle n^{J/\Psi}\rangle$ the average
 number of $J/\Psi$ which are measured generally speaking in a different
 rapidity window.
It turns out that $F\Lb \frac{n}{N}\Rb\,\,\neq\,\,\frac{n}{N}$, but it is
 close to this when the rapidity windows are different. When both rapidity
 windows are the same, $F\Lb \frac{n}{N}\Rb$ shows much steeper dependence
 than 
$\frac{n}{N}$. In Ref.\cite{FEPA} the $J/\Psi$ production is considered as
being
 proportional to the number of collisions, since it comes from  short
 distances, while the production of hadrons is proportional to the number
 of participants (see Ref.\cite{KLN}.) However, in the framework of the 
CGC
 approach,  the $J/\Psi$ production  at high energies is proportional to
 the number of participants\cite{LESI,KHTU,KLNT,DKLMT} as it can be seen
 from \fig{3p}. 

 The main ingredients for describing the experimental data are
 \eq{QCDSOL2}-\eq{QCDSOL4} and \eq{QCDOM6}-\eq{QCDOM7},  as well as
 \eq{TM11}. Using these equation we can re-write \eq{NRDIST02} in the form:
\be  \label{COM2}
\frac{\frac{d \sigma_n^{\rm J/\Psi}}{ d y}}{ \frac{d \sigma^{\rm prod. gl.}_n}{ d y}} \,\,\,=\,\,
\frac{\frac{d \sigma_{\rm incl}^{\rm J/\Psi}}{ d y}}{ \frac{d \sigma^{\rm prod. gl.}_{\rm incl}}{ d y}}
  \frac{1}{1\,\,+\,\,\kappa}\Bigg( \frac{n}{\langle n^{(1)}\rangle} \    \,\,\,+\,\,\, \kappa \frac{n}{\langle n^{(2)}\rangle   }   \, \, \frac{ P^{(2)}_n\Lb N\Rb}{ P^{(1)}_n\Lb N\Rb}\Bigg)  \,\,=\,\,\frac{\frac{d \sigma_{\rm incl}^{\rm J/\Psi}}{ d y}}{ \frac{d \sigma^{\rm prod. gl.}_{\rm incl}}{ d y}}
  \frac{1}{1\,\,+\,\,\kappa}\Bigg( \frac{n}{N} \    \,\,\,+\,\,\, \kappa \frac{n}{2\,N   }   \, \, \frac{ P^{(2)}_n\Lb N\Rb}{ P^{(1)}_n\Lb N\Rb}\Bigg)  
    \ee 
  
  Using \eq{COM2} we can find the experimental observable
 (see Refs.\cite{ALICE1,LSS}  )
 
  \begin{eqnarray}\label{COM3}
 & \displaystyle{\frac{dN_{J/\psi}/dy}{\langle dN_{J/\psi}/dy\rangle}\,\,=\frac{w\left(N_{J/\psi}\right)}{\left\langle w\left(N_{J/\psi}\right)\right\rangle }\,\frac{\left\langle w\left(N_{{\rm ch}}\right)\right\rangle }{w\left(N_{{\rm ch}}\right)}}
 &=  \frac{d\sigma_{J/\psi}\left(y,\,\eta,\,\sqrt{s},\,n\right)/dy}{d\sigma_{J/\psi}\left(y,\,\eta,\,\sqrt{s},\,n=N\right)/dy}/\frac{d\sigma_{{\rm ch}}\left(\eta,\,\sqrt{s},\,n\right)/d\eta}{d\sigma_{{\rm ch}}\left(\eta,\,\sqrt{s},\,n  = N\right)/d\eta}\nonumber \\
 &= \frac{\Bigg( \frac{n}{N} \    \,\,\,+\,\,\, \kappa \frac{n}{2\,N}   \, \, \frac{ P^{(2)}_n\Lb N\Rb}{ P^{(1)}_n\Lb N\Rb}\Bigg)}{  \Bigg(1\   \,\,\,+\,\,\, \h\kappa   \, \, \frac{ P^{(2)}_N\Lb N\Rb}{ P^{(1)}_N\Lb N\Rb}\Bigg)} &\displaystyle{ \xrightarrow{z = \frac{n}{N} \gg1} \frac{ z\,\,+\,\,\frac{\kappa}{4}\,z^2}{1+\frac{\kappa}{4}}}
 \end{eqnarray}

\begin{figure}[ht]
     \begin{center}
     \begin{tabular}{c c c}
     \includegraphics[width=  0.45\textwidth]{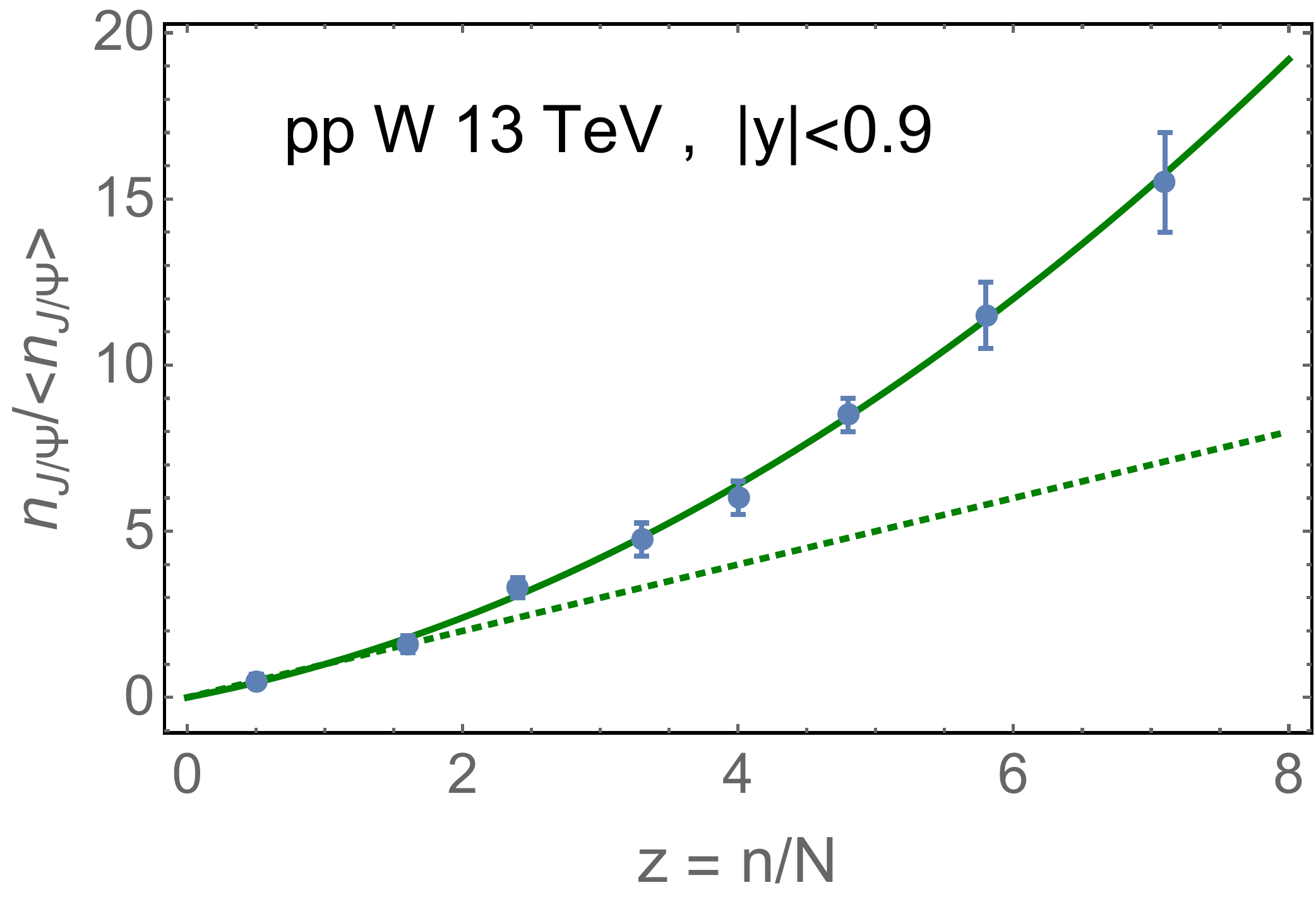} & ~~~~~~~~&    \includegraphics[width=  0.44\textwidth]{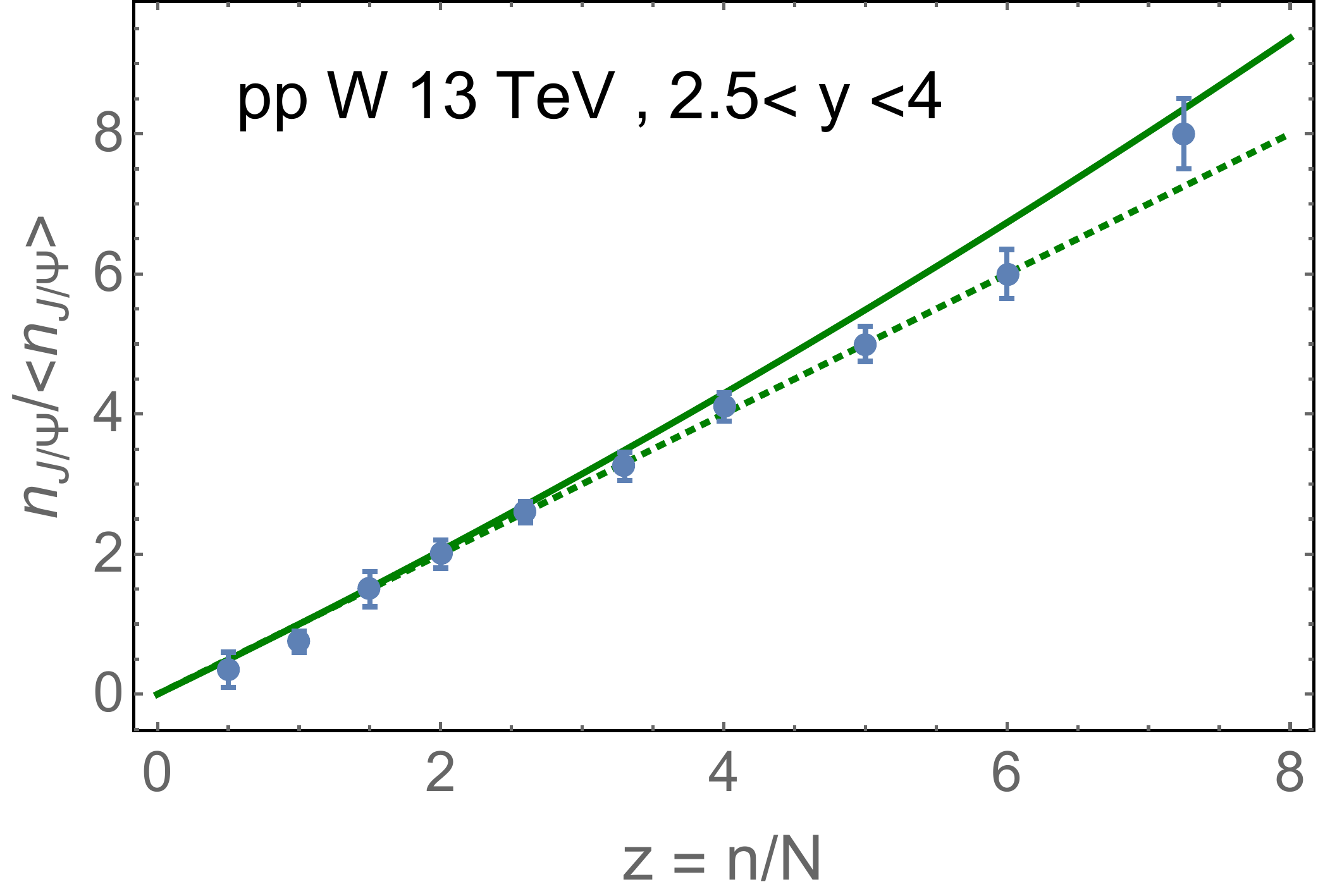} \\
     \fig{psicom}-a & &\fig{psicom}-b\\
     \end{tabular}
         \end{center}  
   \caption{ Comparison \eq{COM3} with the experimental data of the ALICE
 collaboration\cite{ALICE0,ALICE1,ALICE01,ALICE2,ALICE3}. The solid line
 is the estimate of \eq{COM3}, and the dotted line is the linear 
dependence
 which stems from the contribution of \fig{psiinmpsi}-a. \fig{psicom}-a
 shows the production of $J/\Psi$ in central rapidity region, while in
 \fig{psicom}-b  the estimates are shown for \eq{COM3} with 
   $\kappa$ calculated in leading order of perturbative QCD, and with 
$\bas = 0.15$.}
\label{psicom}
   \end{figure}

In \fig{psicom} we compare  the experimental data with 
\eq{COM3}.
  %%%%%%%%%%%%%%%%%%%%%%%%%%%%%%%%%%%%%%%%%%%%%%%%%%%%%%%%%%%%%%%
%%%%%%%%%

One can see that this simple formula provides a  fairly good description 
of  the experimental
 data, for  central production where $\kappa = 1$. However, the 
experimental
 data for forward production of $J/\Psi$\cite{ALICE0,ALICE1,ALICE01,
ALICE2,ALICE3,STAR1,STAR2} show almost  a linear dependance:
 $\frac{n_{J/\Psi}}{\langle n_{J/\Psi}\rangle }\,\,=\,\,z$.
 Indeed, in \eq{COM3} the quadratic term is suppressed, since    
  it is proportional to the value of $\kappa$, which is equal to 
 (see  \fig{psiinmpsi}
     and Ref.\cite{LESI} for the estimates).
\be \label{COM6}
\kappa\,\,\,\,=\,\,\,\Lb \frac{Q^2_s\Lb Y - y\Rb}{Q^2_s\Lb  y\Rb}\Rb^{ \bar{\gamma}}\,\,=\,\,e^{-\,2\, \bar{\gamma} \lambda\,y^*} 
\ee
where $y^*$ is the rapidity of the produced quarkonia in c.m.f.
 In leading order of perturbative QCD , in which we made all our
 previous estimates\cite{KOLEB},
$\bar{\gamma}\,\,=\,\,0.63$ and $\lambda \,=\,\bas \,\frac{\chi\Lb
 \bar{\gamma}\Rb}{\bar{\gamma}}\,\,\approx\,4.8\,\bas$.  As  is
 shown in  \fig{psicom}-b, the estimate in  leading order  describes
 the data quite well.  However,
 we need to remember that the NLO corrections both to $\bar{\gamma}$
 and to  $\lambda$ are large.

~

~
    %  %%%%%%%%%%%%%%%%%%%%%%%%%%%%%%%%%%%%%%%%%%%%%%%%%%
\section{Conclusions}
  %%%%%%%%%%%%%%%%%%%%%%%%%%%%%%%%%%%%%%%%%%%
  
\begin{figure}[ht]
     \begin{center}
     \includegraphics[width=  16cm,height=3.7cm]{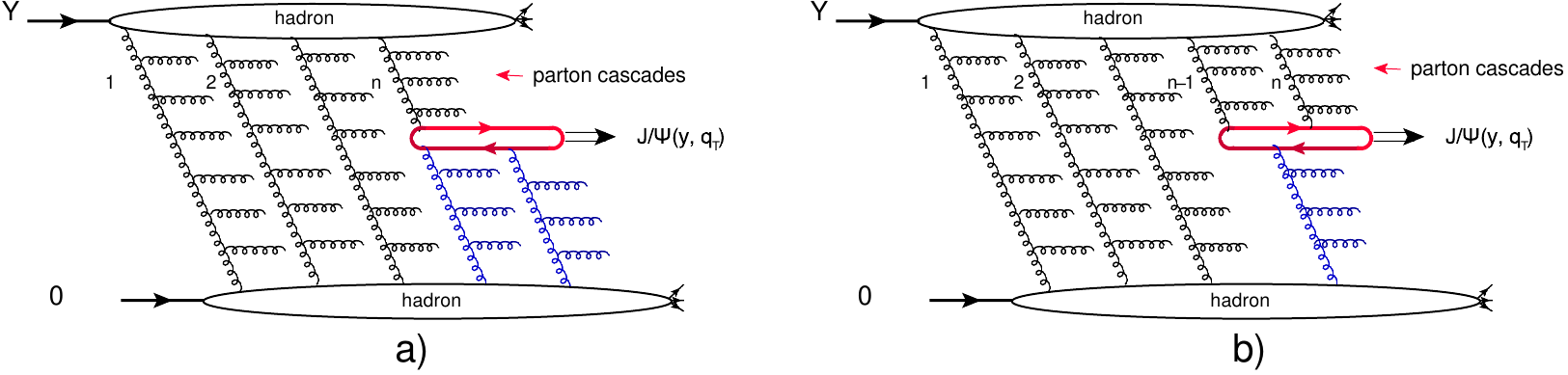} 
     \end{center}  
   \caption{ The production of quarkonia from $n$ parton cascades.}
\label{nlad}
   \end{figure}

  In this paper we re-visited the problem of multiplicity distributions
 in high energy QCD, which we have discussed in Ref.\cite{GOLEMULT} and
 found the distribution of \eq{QCDSOL30}. This distribution provides a 
better
 description of  the experimental data at large multiplicities $n$,  than
 \eq{NDIST}, which has been discussed previously. 
  We also suggest a different approach to the multiplicity dependence
 of quarkonia production. It should be stressed that our approach is
 based on the  three gluons fusion mechanism  of \fig{3p}, and it differs
 from the description of Refs.\cite{MTVW,LSS}, since we did not assume
 the multiplicity dependence of the saturation scale.  In our approach
 we assume, that the production of  $J/\Psi$, which occurs at  rapidity 
$y$,
 and the central production of charged hadrons, stem from the production
 of the same $n$-parton cascades, which are  pictured in \fig{nlad} as
  the production of $n$-gluon ladders.  Solving the QCD cascade
 equation, we found the multiplicity distribution both for the cascade
 of  \fig{nlad}-a (see \eq{QCDOM6} - \eq{QCDOM7}) and for the cascade
 of  \fig{nlad}-b (see \eq{QCDSOL2} - \eq{QCDSOL4}) . 
  
  In \fig{nlad} one can see that $J/\Psi$ can be   produced from each of
 $n$-ladders, leading to the cross section, which is proportional to $n$
 (see \fig{nlad}-a). This mechanism is shown in \fig{psipr}-a. However,
 $J/\Psi$ can be created from merging of two ladders (see \fig{nlad}-b) ,
 which gives a cross section $\propto\,\,n^2$, and corresponds to
 \fig{psipr}-b.  Note, that the production of the hadrons in both
 cases can be found from \eq{TM11}. Taking into account that the average
 number of gluons (hadrons) for the mechanism of \fig{nlad}-b is  two
 time larger than for \fig{nlad}-a, we infer that \fig{nlad} leads to the
 simple \eq{COM3}.

      %%%%%%%%%%%%%%%%%%%%%%%%%%%%%%%%%%%%%%%%%%%%%%%%%%%%%%%%%%%%%%%
%%%%%%% %%%%%%%%%%%%%%%%%%%%%%%%%%%%%%%%%%%%%%%%%%%%%%%%%%

  It should be stressed that this equation is heavily dependent on the
 three gluon fusion mechanism, but does not depend on the details 
of the cross section of quakonia production. In particular, as we
 have mentioned above, we do not use the dependence of the saturation
 scale on the multiplicity of the produced gluons. This means that the 
non-linear dependence of $J/\Psi$-production on multiplicity of charged
 hadrons, can stem from sources other  than the dependence of the cross
 section on the saturation scale. Actually, this statement follows
 directly from the fact that 1+1 RFT generates the non-linear dependence
 on $n$.

 It should be noted that if we assume, that in addition to the three 
gluon
 fusion mechanism we have the production of $J/\Psi$ from color-singlet
 model,  we will not obtain agreement with the description of the 
experimental data in
 \fig{psicom}.

 As an aside, we note in \cite{MTVW,LSS} it was assumed that the 
$J/\Psi$-production results from the inclusive diagrams of   \fig{psipr}-a  and 
\fig{psipr}-b. This is erroneous, as at arbitrary $n$ it is necessary to 
include the production of many partonic showers  as illustrated in   
 \fig{psiinmpsi}-a and \fig{psiinmpsi}-b. The difference to
 the percolation approach\cite{FEPA}, lies in our hypothesis that both
 the production of $J/\Psi$ and the charged pion stem from   short
 distances of the order of $r \, \propto  \,1/Q_s$, and are determined by 
 physics controlled by the CGC effective theory. The gluon jets with  
 transverse momentum $Q_s$, decay into charged pions (see Ref.\cite{LSS} 
for
 details). The non-linear dependence of production of $J/\Psi$  is due to
 the   three Pomeron fusion mechanism.

    In spite of the good description of the experimental data for the
 quarkonia production integrated over the  transverse momenta ($p_T$),
we cannot explain at present, why
 the data at fixed $p_T$ \cite{ALICE0}, shows a steeper dependence
 on $n$ than the integrated data. 
Certainly, this problem will be the main subject of our further
 attempts to understand the multiplicity dependence of quarkonia 
production.

    %  %%%%%%%%%%%%%%%%%%%%%%%%%%%%%%%%%%%%%%%%%%%%%%%%%%
\section{Acknowledgements}
  %%%%%%%%%%%%%%%%%%%%%%%%%%%%%%%%%%%%%%%%%%% 
   We thank our colleagues at Tel Aviv university and UTFSM for
 encouraging discussions.  This research was supported  by 
 ANID PIA/APOYO AFB180002 (Chile) and  Fondecyt (Chile) grants  
 1180118.

\end{document}